\documentclass[11pt,]{article}
\pdfoutput=1
\usepackage{amsmath,amsfonts,amssymb,stmaryrd}
\usepackage[nosort]{cite}
\usepackage[hyperfootnotes=false]{hyperref}
\usepackage{color}
\usepackage{graphicx}
\usepackage{titlesec}
\usepackage{ulem}
\usepackage{color}
\usepackage[font=small,labelfont=bf]{caption}
\usepackage[flushmargin]{footmisc}
\usepackage{bigfoot}
\DeclareNewFootnote[para]{A}
\usepackage{tabularx}

\DeclareSymbolFont{euleroperators}{U}{\sfdefault}{m}{n}
\SetSymbolFont{euleroperators}{bold}{U}{\sfdefault}{b}{n}

\makeatletter
\renewcommand{\operator@font}{\mathgroup\symeuleroperators}
\makeatother

\newif\ifaffiliation
\affiliationtrue

\titlespacing*\section{0pt}{16pt plus 4pt minus 6pt}{4pt plus 2pt minus 2pt}
\titlespacing*\subsection{0pt}{4pt plus 4pt minus 2pt}{2pt plus 2pt minus 2pt}
\titleformat{\section}{\fontsize{14}{15}\bfseries\boldmath}{\thesection}{1em}{}[\fontsize{14}{14}\mdseries\@authors]
\titleformat{\subsection}{\normalfont\fontsize{11}{12} \bfseries\boldmath}{}{0pt}{}
\numberwithin{equation}{section}

\newcommand\authors[1]{\def\@authors{#1}}

\DeclareMathOperator{\Tr}{Tr}

\DeclareMathOperator{\sTr}{sTr}

\DeclareMathOperator{\diag}{diag}

\def\bR {\mathbb{R}}

\def\bZ {\mathbb{Z}}

\newcommand{\cA}{{\mathcal A}}
\newcommand{\cL}{{\mathcal L}}

\newcommand{\cR}{{\mathcal R}}

\newcommand{\cN}{{\mathcal N}}

\newcommand{\su}{\mathfrak{s}\mathfrak{u}}

\newcommand{\osp}{{\mathfrak{osp}}}

\newcommand{\be}{\begin{equation}}
\newcommand{\ee}{\end{equation}}

\newcommand{\folder}{.}

\begin{document}
\thispagestyle{empty}
\noindent{\bf\LARGE Roadmap on Wilson loops in 3d Chern-Simons-matter theories}
\vskip-4mm
\par\noindent\rule[-2mm]{0.7\textwidth}{0.5pt}
\\[-8mm]
\begin{flushleft}
{Editors:}
{\bf Nadav~Drukker}\footnoteA{\href{mailto:nadav.drukker@gmail.com}{nadav.drukker@gmail.com}}\textsuperscript{,a}
and 
{\bf Diego~Trancanelli}\footnoteA{\href{mailto:dtrancan@gmail.com}{dtrancan@gmail.com}}\textsuperscript{,b}
\\
{Contributors:}
{\bf Lorenzo~Bianchi}\footnoteA{\href{mailto:lorenzo.bianchi.ph@gmail.com}{lorenzo.bianchi.ph@gmail.com}}\textsuperscript{,c},
{\bf Marco~S.~Bianchi}\footnoteA{\href{mailto: marcowhites84@gmail.com}{marcowhites84@gmail.com}}\textsuperscript{,d},
{\bf Diego~H. Correa}\footnoteA{\href{mailto:diegocorrea@gmail.com}{diegocorrea@gmail.com}}\textsuperscript{,e},
{\bf Valentina~Forini}\footnoteA{\href{mailto:valentina.forini@city.ac.uk}{valentina.forini@city.ac.uk}}\textsuperscript{,f},
{\bf Luca~Griguolo}\footnoteA{\href{mailto:griguolo1965@gmail.com}{griguolo1965@gmail.com}}\textsuperscript{,g},
{\bf Matias~Leoni}\footnoteA{\href{mailto:leoni@df.uba.ar}{leoni@df.uba.ar}}\textsuperscript{,h},
{\bf Fedor~Levkovich-Maslyuk}\footnoteA{\href{mailto:fedor.levkovich@gmail.com}{fedor.levkovich@gmail.com}}\textsuperscript{,i},
{\bf Gabriel~Nagaoka}\footnoteA{\href{mailto:g.n.nagaoka@gmail.com}{g.n.nagaoka@gmail.com}}\textsuperscript{,j},
{\bf Silvia~Penati}\footnoteA{\href{mailto:silvia.penati@mib.infn.it}{silvia.penati@mib.infn.it}}\textsuperscript{,k},
{\bf Michelangelo~Preti}\footnoteA{\href{mailto:michelangelo.preti@gmail.com}{michelangelo.preti@gmail.com}}\textsuperscript{,l},
{\bf Malte~Probst}\footnoteA{\href{mailto:mltprbst@gmail.com}{mltprbst@gmail.com}}\textsuperscript{,a},
{\bf Pavel~Putrov}\footnoteA{\href{mailto:putrov@ictp.it}{putrov@ictp.it}}\textsuperscript{,m},
{\bf Domenico~Seminara}\footnoteA{\href{mailto:domesemi@gmail.com}{domesemi@gmail.com}}\textsuperscript{,n},
{\bf Guillermo~A.~Silva}\footnoteA{\href{mailto:guilleasilva@gmail.com}{guilleasilva@gmail.com}}\textsuperscript{,e},
{\bf Marcia~Tenser}\footnoteA{\href{mailto:marciatenser@gmail.com}{marciatenser@gmail.com}}\textsuperscript{,j},
{\bf Maxime~Tr\'epanier}\footnoteA{\href{mailto:trepanier.maxime@gmail.com}{trepanier.maxime@gmail.com}}\textsuperscript{,a},
{\bf Edoardo~Vescovi}\footnoteA{\href{mailto: e.vescovi@imperial.ac.uk}{e.vescovi@imperial.ac.uk}}\textsuperscript{,o},
{\bf Itamar~Yaakov}\footnoteA{\href{mailto:itamar.yaakov@gmail.com}{itamar.yaakov@gmail.com}}\textsuperscript{,g}
and
{\bf Jiaju~Zhang}\footnoteA{\href{mailto:jzhang@sissa.it}{jzhang@sissa.it}}\textsuperscript{,p}
\end{flushleft}
\bgroup
\setlength\tabcolsep{0pt}
\begin{tabularx}{\textwidth}{rX}
\textsuperscript{a}&
Department of Mathematics, King's College London, The Strand, London WC2R
2LS, United Kingdom\\
\textsuperscript{b}&
Dipartimento di Scienze Fisiche, Informatiche e Matematiche, Universit\`a di Modena e Reggio Emilia, via Campi 213/A, 41125 Modena \& INFN Sezione di Bologna, via Irnerio 46, 40126 Bologna, Italy\\
\textsuperscript{c}&
Center for Research in String Theory, School of Physics and Astronomy, 
Queen Mary University of London, Mile End Road, London E1 4NS, United Kingdom\\
\textsuperscript{d}&
Instituto de Ciencias F\'{\i}sicas y Matem\'aticas, Universidad
Austral de Chile, 
Valdivia, Chile\\
\textsuperscript{e}&
Instituto de F\'isica de La Plata (IFLP) - CONICET \& Departamento de F\'isica, Facultad de Ciencias Exactas, Universidad Nacional de La Plata C.C. 67, 1900 La Plata, Argentina
\\
\textsuperscript{f}&
Department of Mathematics, City, University of London, Northampton Square, EC1V 0HB London, UK\\
&\& Institut f\"ur Physik, Humboldt-Universit\"at zu Berlin, Zum Gro{\ss}en Windkanal 6, 12489 Berlin, Germany\\
\textsuperscript{g}&
Dipartimento di Scienze Matematiche Fisiche e Informatiche, Universit\`a di Parma \& INFN Gruppo
Collegato di Parma, Parco Area delle Scienze 7/A, 43124 Parma, Italy
\\
\textsuperscript{h}&
Departamento de F\'isica, Universidad de Buenos Aires \& IFIBA - CONICET Ciudad Universitaria, pabell\'on 1 (1428) Buenos Aires, Argentina
\\
\textsuperscript{i}&
Departement de Physique, \'Ecole Normale Sup\'erieure / PSL Research University, CNRS, 24 rue
Lhomond, 75005 Paris, France \& Institute for Information Transmission Problems, Moscow 127994, Russia
\\
\textsuperscript{j}&
Instituto de F\'isica, Universidade de S\~ao Paulo, 05314-970 S\~ao Paulo, Brazil
\\
\textsuperscript{k}&
Universit\`a degli Studi di Milano-Bicocca \& INFN Sezione di Milano-Bicocca, Piazza della
Scienza 3, 20161 Milano, Italy
\\
\textsuperscript{l}&
Nordita, KTH Royal Institute of Technology \& Stockholm University,
Roslagstullsbacken 23, SE-10691 Stockholm, Sweden
\\
\textsuperscript{m}& 
ICTP, Strada Costiera 11, 34151 Trieste, Italy
\\
\textsuperscript{n}&
Dipartimento di Fisica e Astronomia, Universit\`a di Firenze \& INFN Sezione 
di Firenze, via G. Sansone~1, 50019 Sesto Fiorentino, Italy
\\
\textsuperscript{o}&
The Blackett Laboratory, Imperial College London, London SW7 2AZ, United Kingdom
\\
\textsuperscript{p}&
SISSA \& INFN Sezione di Trieste, Via Bonomea 265, 34136 Trieste, Italy
\end{tabularx}
\egroup

\vspace*{.5cm}
\vspace*{.5cm}
\vspace*{.5cm}
\vspace*{.5cm}
\vspace*{.5cm}
\eject
\thispagestyle{empty}
\noindent{\bf Abstract}\\[2mm]
\begin{minipage}{0.9\textwidth}
This is a compact review of recent results on supersymmetric Wilson loops in ABJ(M) and related theories. It aims to be a quick introduction to the state of the art in the field and a discussion of open problems. It is divided into  
short chapters devoted to different questions and techniques. Some new results,  perspectives and  speculations are also presented. We hope this might serve as a baseline for further studies of this topic.\\ \emph{Prepared for submission to J. Phys. A.}
\end{minipage}

\setcounter{page}{0}
\vspace*{1cm}
\authors{}
\setcounter{tocdepth}{1}
{\boldmath\tableofcontents}
\newpage

\ifaffiliation
\authors{Nadav Drukker, King's College London\\*
Diego Trancanelli, UniMORE and INFN Bologna}
\else
\authors{Nadav Drukker and Diego Trancanelli}
\fi
\section{Overview}
\label{chapter:1}

\subsection{Genesis}
This roadmap article grew out of a fortunate confluence. Recently, one of us organized a workshop at the University of Modena and Reggio Emilia,\footnote{The {\it Mini-workshop on supersymmetric Wilson loops and related topics} took place on May 15/16, 2019. For details see {\tt https://agenda.infn.it/event/19090/}.} which was attended by many researchers who have worked on Wilson loops in three-dimensional supersymmetric Chern-Simons-matter theories. The number of participants and the talks they presented were a powerful illustration of the breadth of work on this topic. Moreover, in the run-up to the conference, together with the authors of Chapter~\ref{chapter:2} we realized, as part of ongoing research, that there is a simpler formulation of the 1/2 BPS Wilson loop \cite{Drukker:2009hy} of ABJ(M) theory \cite{Aharony:2008ug,Aharony:2008gk}, which is at the source of much of the research presented in the workshop. We thought that this new formulation could prompt a refreshed look at this research area, encouraging novel attempts to address the many questions which are still left unanswered or only partially understood.

We have therefore decided to put the status of this research area on paper. We suggested to the workshop participants (and several others who could not attend) to contribute to this roadmap and each volunteered a topic that they wanted to cover. They were instructed to follow the new formulation in presenting the salient results, and also focus on the open questions. We hope that this review and this new approach will be beneficial both to the people already working on this area, by nudging them a bit from the comfort of their preferred way of thinking, and to people from the outside who would like to know the state of the art and possibly contribute to it.

In the rest of this introduction we discuss overall themes and questions that permeate throughout the chapters of this review.

\subsection{State of the art}

Wilson loops in supersymmetric gauge theories are particularly interesting observables to study, both because of the possibility of computing them exactly in some cases and because of their relevance in the AdS/CFT correspondence, where they give rise to a rich dictionary between gauge theory and string theory quantities. While supersymmetric Wilson loops in four-dimensional ${\cal N}=4$ super Yang-Mills (SYM) theory have been studied extensively and are well understood for two decades now, their three-dimensional counterparts have a shorter, but arguably more interesting, life. 

After the original formulation of the ABJM theory in 2008, the so-called {\it bosonic} Wilson loop was readily constructed \cite{Berenstein:2008dc,Drukker:2008zx,Chen:2008bp,Rey:2008bh}. This operator only couples to the gauge field and the scalar fields of the theory and turns out to preserve 1/6 of the supersymmetries of the theory. It took some time to construct a more supersymmetric operator, preserving half of the supersymmetries \cite{Drukker:2009hy}. Its formulation mixes the two gauge groups and includes coupling to the fermionic fields as well. Moreover, the supersymmetry variation of this loop does not vanish locally, but it is rather a total derivative along the loop. 

To preserve half of the supersymmetries, the loop has to be a straight line or a circle, although one can generalize the construction such that loops with more general shapes still preserve global supersymmetry \cite{Cardinali:2012ru}, as explained in Chapter~\ref{chapter:7}. Rather awkward features of these original constructions were the explicit dependence of the fermionic coupling on the contour parametrization, as well as the need of introducing a global twist matrix to enforce the correct periodicity conditions. The new formulation presented in Chapter~\ref{chapter:2} remedies these shortcomings in a rather elegant way.

The evaluation of the circular Wilson loops can quite easily be represented by a matrix model, after using localization techniques and a cohomological equivalence argument, as reviewed in Chapter~\ref{chapter:3}. The solution of this matrix model is sketched in Chapter~\ref{chapter:4}. After arduous work to develop perturbative methods in ABJM theory, the subject of Chapter~\ref{chapter:5}, the matrix model solution was matched to the perturbative calculation. The issue of framing of the loops turned out to be more subtle than in pure topological Chern-Simons, as explained in Chapter~\ref{chapter:6}. A proposal for a matrix model for the latitude Wilson loop of Chapter~\ref{chapter:7} is reviewed in Chapter~\ref{chapter:8}.

The perturbative tools allowed also to calculate cusped Wilson loops and a variety of Bremsstrahlung 
functions (see Chapters~\ref{chapter:10} and \ref{chapter:11}). In Chapter~\ref{chapter:13}, a conjecture for the exact form of the function $h(\lambda)$ is motivated from relating it to the matrix model for the 1/6 BPS Wilson loop. This function enters all integrability results, such as the giant magnon dispersion relation discussed in Chapter~\ref{chapter:14}.

A (perhaps unnecessarily) long time has passed until these constructions were generalized to theories 
with non-maximal supersymmetry. A surprising feature is the large parameter space of circular BPS Wilson 
loops with fermionic couplings, including previously (and for the most part also subsequently) ignored 1/6 
BPS Wilson loops in the ABJM theory itself. These exist for theories with $\cN\geq2$, but to keep the discussion relatively brief and concrete, we focus on the generalization to theories with $\cN=4$, as reviewed in Chapter~\ref{chapter:9}.

A topic that has seen very small progress since the first days of this topic is the holographic duals of these Wilson loops. Clearly the 1/2 BPS loop is dual to a fundamental string in $AdS_4\times{\mathbb C \mathbb P}^3$, or an M2-brane in $AdS_4\times S^7/{\mathbb Z}_k$. A full understanding of the less supersymmetric loops as well as the analogs of ``giant Wilson loops'' as the D3-branes and D5-branes in $AdS_5\times S^5$, or ``bubbling geometries'', are thus far lacking. Still some nice work on the cases that are understood has been achieved, and it is presented in Chapters~\ref{chapter:15} and \ref{chapter:12}.

Another topic that one would hope could tie in to this discussion, but was never elucidated, is how to implement integrability for cusped Wilson loops in ABJM theory. The example of ${\cal N}=4$ super Yang-Mills (SYM) suggests an open version of the integrable models describing the spectral problem in ABJM. This requires finding the appropriate boundary conditions (or boundary reflection matrices) for this problem. Though this has been studied by several groups, these works were never completed and nothing has ever been published on these attempts. Chapter~\ref{chapter:14} is the first such attempt to present this question in print and we hope that it will lead to progress. 

\subsection{Future directions}
Each of the chapters presents some open problems, let us only highlight three here.

\begin{itemize}
\item
There is an abundance of different operators sharing the same symmetries. An important open question is to understand their moduli space, whether it is lifted or modified at the quantum level and what are the holographic duals of these operators. 

\item
The new formulation of the fermionic Wilson loops introduced in this roadmap raises several questions and possibilities. Is there an interpretation of the constant pieces in the connection in~\eqref{2:superconnection} as a background field? Is there a simpler way to implement localization for these loops without using their cohomological equivalence to the 1/6 BPS bosonic loop? 

\item
It would be interesting to extend the analyses of Bremsstrahlung functions to other operators, including the fermionic 1/6 loops and loops in ${\cal N}=4$ theories.

\end{itemize}

\subsection{Acknowledgements}
We are grateful to O. Corradini for his valuable help with the organisation of the event at the University of Modena and Reggio Emilia, from which this review has originated. We received financial support through a joint King's College-FAPESP grant. ND is supported by the STFC grant ST/P000258/1. DT is supported in part by the INFN grant Gauge and String Theory (GAST).

\ifaffiliation
\subsection{Full affiliation}
Nadav Drukker,
Department of Mathematics, King's College London, 
The Strand, WC2R 2LS London, United Kingdom
\\
Diego Trancanelli,
Dipartimento di Scienze Fisiche, Informatiche e Matematiche, Universit\`a di Modena e Reggio Emilia, via Campi 213/A, 41125 Modena \& INFN Sezione di Bologna, via Irnerio 46, 40126 Bologna, Italy
\fi
\ifaffiliation
\authors{Gabriel Nagaoka and Marcia Tenser, University of S\~ao Paulo
\\*
Malte Probst and Maxime Tr\'epanier, King's College London}
\else
\authors{Gabriel Nagaoka, Malte Probst, Marcia Tenser and Maxime Tr\'epanier}
\fi
\section{1/2 BPS and 1/6 BPS circular Wilson loops}
\label{chapter:2}

\subsection*{Background}
Circular Wilson loops in ABJ(M) generally fall in two categories: the
bosonic loops coupling to vector and scalar
fields~\cite{Drukker:2008zx,Chen:2008bp,Rey:2008bh}, and the fermionic loops
which also couple to fermions and notably include a family of 1/6 BPS
loops~\cite{Ouyang:2015iza} and the 1/2 BPS loop~\cite{Drukker:2009hy}. While
the discovery of these operators opens many directions of research, key aspects
of their construction remain riddled with intricacies. In particular, these
loops have hitherto not been written in a manifestly gauge and reparametrisation
invariant way.

In this chapter we clarify some of these issues by giving a gauge equivalent formulation of the 
same operators. We find that, in this new language, the generic 1/6 BPS fermionic operator can 
be written naturally as a deformation of a bosonic loop, shedding new light on how these 
loops preserve supersymmetry and their moduli space.

\subsection*{Bosonic loops}

The simplest 1/6 BPS loops of ABJ(M) are bosonic. These can be 
understood as the 1/2 BPS loops of $\mathcal{N} = 2$ theories, which were
found in~\cite{Gaiotto:2007qi} and take the
form\footnote{In accordance with~\cite{Drukker:2009hy} we take the path
ordering to be left-to-right, so that the covariant derivative
is $\partial_\mu + i \left[ A_\mu, \cdot \right]$.}
\begin{equation}
\label{2:bosonic}
  W^\text{bos}_R = \Tr_R \mathcal{P} \exp
  \left(i \oint
  \cA_\text{bos} d \tau\right), \qquad
  \cA_\text{bos} = A_{\mu} \dot{x}^\mu
  - i \sigma |\dot{x}|,
\end{equation}
where the loop is taken over a circle and
$\sigma$ is the auxiliary field in the $\mathcal{N} = 2$ vector
multiplet. In the $\mathcal{N} = 6$ theory they can be defined independently
for both gauge groups $U(N_1)$ and
$U(N_2)$~\cite{Drukker:2008zx,Chen:2008bp,Rey:2008bh}, for which the auxiliary fields
are fixed to $\sigma^{(1)} = 2\pi k^{-1} M_J^I C_I \bar{C}^J$ with 
$M= \diag{(-1,-1,+1,+1)}$ up to equivalent choices, with $C_I$, $\bar C^J$ bifundamental scalars, 
and similarly for $\sigma^{(2)}$. 
As is evident from the structure of $M$, the loop has residual 
$SU(2) \times SU(2)$ R-symmetry. These Wilson loops preserve four supercharges parametrised by
anticommuting $\theta_{12}^\pm$ and $\bar{\theta}^{12}_\pm$,
accompanied by the special supersymmetries fixed to\footnote{These conditions can 
equivalently be stated in terms of $\theta_{34} = -\bar{\theta}^{12}$, see the end of this chapter.}
\begin{align}
  \label{2:bps}
  \epsilon_{12} = - i \theta_{12} \frac{\gamma_3}{|x|}, \qquad
  \bar{\epsilon}^{12} = - i \frac{\gamma_3}{|x|} \bar{\theta}^{12},\qquad 
  \gamma_3=\text{diag}(1,-1).
\end{align}

\subsection*{Fermionic loops}

A wider class of 1/6 BPS operators preserving the same supercharges can be
constructed from the two bosonic loops by allowing for nonzero coupling to the
bifundamental fermions. These fermionic loops take the form
\begin{equation}
\label{2:loop}
W^\text{fer}_\cR = (-i)^{|\cR|} \sTr_\cR \mathcal{P} \exp
\left(i \oint\mathcal{L}(x^\mu,\dot x^\mu) d \tau\right),
\end{equation}
where $\cR$ indicates a representation of $U(N_1|N_2)$ whose Young diagram has 
weight $|\cR|$, the
superconnection $\cL$ is given by a deformation of the composite bosonic
connection,
\begin{align}
\label{2:bosonicdeformation}
\cL = \cL_\text{bos} + \Delta \cL, 
\qquad
    \cL_\text{bos} = \begin{pmatrix}
    \mathcal{A}_\text{bos}^{(1)} & 0\\
    0 & \mathcal{A}_\text{bos}^{(2)}\\
    \end{pmatrix} + \frac{|\dot{x}|}{4 |x|} \sigma_3,
\end{align}
and $\Delta \cL$ may be block off-diagonal.
For $\Delta \cL = 0$, the constant term $\sigma_3 =
\text{diag}(+\mathbb{I}_{N_1}, - \mathbb{I}_{N_2})$ can be
exponentiated and we recover the sum of the usual trace of bosonic connections.

In order for~\eqref{2:loop} to preserve supersymmetry, we require the
superconnection to transform under the preserved supercharges as 
$\delta\cL = \mathfrak{D}_\tau \mathcal{G} \equiv \partial_\tau \mathcal{G}+i[\cL,\mathcal{G}]$ for some 
$\mathcal{G}$~\cite{Drukker:2009hy,Lee:2010hk}. 
This relaxed notion of supersymmetry ensures that the variation takes the form
of a supergauge transformation, under which the loop~\eqref{2:loop} is
invariant. Consider then a deformation
\begin{align}
  \label{2:deformation}
  \Delta \cL = i |\dot x | \sigma_3 \left( \delta_+ \mathcal{G} + i
  \mathcal{G}^2 \right),
\end{align}
where $\delta_+$ is parametrised by $\theta^+_{12}$, $\bar\theta_+^{12}$ (and 
$\epsilon^+_{12}$, $\bar\epsilon_+^{12}$ are given by \eqref{2:bps}).
The variation of $\Delta \cL$ with respect to $\delta_+$ assembles into a total 
derivative as required if 
\begin{align}
  \label{2:susy}
  \delta_+^2 \mathcal{G} = - i \sigma_3 \left( \partial_\tau \mathcal{G} +
  [\cL_\text{bos}, \mathcal{G}] \right), \qquad
  \theta_{12}^+ \bar{\theta}^{12}_+ = \frac{1}{4},
\end{align}
which is satisfied for $\cal G$ comprised of 
$C_1$, $\bar C^1$ $C_2$, $\bar C^2$, breaking one $SU(2)$ of the residual
R-symmetry (here, $\alpha, \bar\alpha \in \mathbb{C}^2$ are taken to be
Grassmann odd and $i,j=1,2$)
\begin{align}
  \mathcal{G} = \sqrt{\frac{\pi i}{2 k}} \begin{pmatrix}
    0 & \bar{\alpha}^i C_i\\
   -\alpha_i \bar{C}^i & 0\\
  \end{pmatrix}.
\end{align}
Using~\eqref{2:susy} one can show easily that $\delta_+ \cL = 
\mathfrak{D}_\tau \mathcal{G}$. Invariance under $\delta_-$ (parametrised by the
remaining parameters $\theta_{12}^-,$ ${\bar\theta}^{12}_-$) is ensured because $\delta_- \mathcal{G}$ is
related to $\delta_+ \mathcal{G}$ by a gauge transformation, so that $\delta_-
\cL$ also takes the form of a total derivative.

The resulting family of 1/6 BPS loops is then parametrised by $\alpha,
\bar{\alpha}$ and can be written explicitly as
\begin{gather}
\label{2:superconnection}
\mathcal{L}= \begin{pmatrix}
  \mathcal{A}^{(1)} &&
  \sqrt{-\frac{4\pi i }{k}}|\dot x| \eta_i \bar{\psi}^i\\
  \sqrt{-\frac{4\pi i}{k}} |\dot x|\psi_i \bar{\eta}^i &&
  \mathcal{A}^{(2)}\\
  \end{pmatrix},\qquad 
  \begin{matrix}
  \mathcal{A}^{(1)} = \cA^{(1)}_\text{bos}
  - \frac{2\pi i}{k} |\dot{x}| \Delta M_j^i C_i \bar{C}^j
  +\frac{|\dot{x}|}{4|x|}\,,\\[3pt]
  \mathcal{A}^{(2)} = \cA^{(2)}_\text{bos}
  - \frac{2\pi i}{k} |\dot{x}| \Delta M_j^i \bar{C}^j C_i
  - \frac{|\dot{x}|}{4|x|}\,,
  \end{matrix}\\
  \label{2:eta}
  \eta_i =
  2 \sqrt{2} \bar{\alpha}^j \theta_{ij}^+ \Pi_+, \qquad
  \bar{\eta}^i =
  2 \sqrt{2} \Pi_+ \bar{\theta}^{ij}_+ \alpha_j, \qquad
  \Delta M^i_j = 2 \bar{\alpha}^i \alpha_j, \qquad
  \Pi_\pm \equiv \frac{1}{2}\left(1\pm \frac{\dot{x}^\mu \gamma_\mu}{|\dot
  x|}\right).
\end{gather}

We note that~\eqref{2:superconnection} is related to the operators
of~\cite{Drukker:2009hy,Ouyang:2015iza} by a gauge transformation parametrised
by $\Lambda = (\pi - 2 \phi)/8 \cdot \sigma_3$,
where $0 < \phi < 2
\pi$ is the polar angle and $\pi/8$ accounts for different conventions for
$\eta, \bar{\eta}$. The fields transform as
\begin{equation}
  \label{2:gaugetransformation}
  \mathcal{A}^{(1)} \rightarrow \mathcal{A}^{(1)} -  \frac{|\dot{x}|}{4 |x|}, \qquad 
  \mathcal{A}^{(2)} \rightarrow \mathcal{A}^{(2)} +  \frac{|\dot{x}|}{4 |x|}, \qquad 
  \psi \rightarrow \sqrt{-i} e^{-i \phi/2} \psi, \qquad \bar\psi\rightarrow
  \sqrt{i} e^{i\phi/2}\bar\psi\,,
\end{equation}
where the right-hand side reproduces the original formulation. The discontinuity of 
$\Lambda$ at $2\pi$ yields a delta function term which can be integrated to
exchange the supertrace for a trace.

We stress that in contrast to previous formulations, \eqref{2:loop} is 
manifestly reparametrisation invariant. It is also gauge invariant without the need
for an additional twist matrix (see for
instance~\cite{Cardinali:2012ru}), since the couplings $\eta, \bar{\eta}$ and $M
+ \Delta M$ are periodic by construction. This comes, of course, at the expense
of introducing a constant piece in the connection $\cL_\text{bos}$, whose
physical interpretation remains unclear.

We obtain the moduli space of 1/6 BPS deformations~\eqref{2:deformation} 
by noting that any rescaling $\alpha$ and $\bar\alpha$ such that their product 
$\Delta M$ is unmodified can be absorbed by a gauge transformation. The resulting 
manifold is the space of singular complex matrices, which is the conifold. 
This construction matches Class I of~\cite{Ouyang:2015iza}, while Class II is obtained 
by breaking the other $SU(2)$, {\it i.e.}\ coupling to $C_3$, $\bar C^3$, $C_4$ and $\bar C^4$ in $\cal G$. 
These two branches intersect at the origin singularity $\cL_\text{bos}$. 

At particular points where $\Delta M$ has eigenvalues $2$ and 0, 
the full matrix $M+\Delta M$ has enhanced $SU(3)$ symmetry. 
It is easy to see that commuting the 4 preserved supercharges
with this $SU(3)$ symmetry gives rise to 12 supercharges, so these 
operators are 1/2 BPS, and we recover the loops of~\cite{Drukker:2009hy}, 
as may be checked explicitly by performing the gauge transformation used above.

It would be interesting to check whether~\eqref{2:deformation} covers the
full moduli space of 1/6 BPS loops and to further investigate the geometry of 
these moduli. In particular, it would be interesting to study deformations
around a generic point of the moduli space and to understand the constant
diagonal term in $\cL_\text{bos}$ as a geometric feature of that space.

\subsection*{Conventions and notations}

We mostly adopt the conventions of~\cite{Cardinali:2012ru} and denote the gauge
group of ABJ(M) theory as $U(N_1) \times U(N_2)$. In addition to the
gauge fields $A^{(1)}$ and $A^{(2)}$ transforming in the adjoint of their
respective gauge group, the theory contains scalars $C_{I}$ and $\bar{C}^I$ and
fermions $\psi^\alpha_I$ and $\bar\psi^I_\alpha$ in the bifundamental, such that
$C \bar{C}$ and $\bar{\psi} \psi$ ($\bar{C} C$ and $\psi \bar{\psi}$) transform in
the adjoint of $U(N_1)$ ($U(N_2)$), with the R-symmetry index $I$ transforming
in the fundamental of $SU(4)$. These fields assemble in a single
supermultiplet satisfying
\begin{equation}
  \label{2:susyrules}
  \begin{aligned}
     \delta A^{(1)}_{\mu} &=
    - \frac{4\pi i}{k}C_I \psi_J^\alpha {(\gamma_\mu)_\alpha}^\beta \bar\Theta^{IJ}_\beta
    +\frac{4\pi i}{k} \Theta_{IJ}^\alpha{(\gamma_\mu)_\alpha}^\beta
    \bar\psi_\beta^I \bar{C}^J\,, \\
    \delta A^{(2)}_{\mu} &=
    \frac{4\pi i}{k} \psi_I^\alpha C_J {(\gamma_\mu)_\alpha}^\beta \bar\Theta^{IJ}_\beta
    - \frac{4\pi i}{k}\Theta_{IJ}^\alpha {(\gamma_\mu)_\alpha}^\beta
    \bar{C}^I\bar\psi_\beta^J\,, \\
    \delta\bar\psi_\beta^I &=
    2i {(\gamma^\mu)_\beta}^\alpha \bar\Theta^{IJ}_\alpha D_\mu C_J
    +\frac{16\pi i}{k}\bar\Theta_\beta^{J[I} C_{[J} \bar{C}^{K]} C_{K]}
    -2i 
    \bar\epsilon_\beta^{IJ} C_J\,,\\
    \delta\psi_I^\beta &=
    -2 i \Theta_{IJ}^\alpha {(\gamma^\mu)_\alpha}^\beta D_\mu \bar{C}^J
    -\frac{16\pi i}{k}\Theta_{J[I}^\beta \bar{C}^{[J} C_{K]} \bar{C}^{K]}
    -2i 
    \epsilon_{IJ}^\beta \bar{C}^J\,, \\
    \delta C_I &=
    2 \Theta_{IJ}^\alpha \bar\psi_\alpha^J\,, \\ 
    \delta\bar{C}^I &=
    - 2 \psi_J^\alpha \bar\Theta^{JI}_\alpha\, ,
\end{aligned}
\end{equation}
for a (Euclidean) superconformal transformation parametrised by
$\Theta_{IJ} = \theta_{IJ} + \epsilon_{IJ} (x \cdot \gamma)$ and
$\bar{\Theta} = \bar{\theta}^{IJ} - (x \cdot \gamma) \bar{\epsilon}^{IJ}$.
The parameters are related by
$\bar{\theta}^{IJ}_\alpha = -\frac{1}{2} \varepsilon^{IJKL} \theta_{KL}^\beta \varepsilon_{\beta \alpha}$
(likewise $\bar\epsilon_\alpha^{IJ}$), but unlike in Minkowski space there is no
reality condition ({\it i.e.}\ $\bar{\theta} \neq \theta^\dagger$).
Omitted spinor indices $\alpha = \pm$ follow the NW-SE summation convention.
A review of the theory in these conventions along with
an action can be found in~\cite{Bianchi:2013rma}.

\subsection{Acknowledgements}
We would like to thank N. Drukker and D. Trancanelli for their support, enlightening discussions
and comments on the manuscript.
M. Tr\'epanier acknowledges the support of the Natural Sciences and Engineering Research Council 
of Canada (NSERC). M. Tenser and G. Nagaoka acknowledge the support of the Conselho Nacional de Desenvolvimento
Cient\'ifico e Tecnol\'ogico (CNPq).

\ifaffiliation
\subsection{Full affiliation}
Gabriel Nagaoka and Marcia Tenser, Institute of Physics, University of S\~ao Paulo, S\~ao Paulo, 05314-970, Brazil\\
Malte Probst and Maxime Tr\'epanier, Department of Mathematics, King's College London, London, WC2R 2LS, United Kingdom
\fi

\authors{Itamar Yaakov\ifaffiliation, Universit\`a di Parma and INFN Parma\fi}
\section{Localization of BPS Wilson loops and cohomological equivalence}
\label{chapter:3}

\subsection{Background}

It is possible to evaluate the expectation value of the BPS Wilson loop $W^\text{bos}_{R}$ 
\eqref{2:bosonic} of Chapter~\ref{chapter:2} exactly, even at strong coupling. We begin by making a Weyl transformation to an $S^3$ of radius $r$. It is straightforward to formulate a theory on curved space
while preserving supersymmetry provided the space admits parallel (Killing) spinors,
{\it i.e.}\ spinors satisfying $\nabla_{\mu}\epsilon=0$. While these do
not exist on $S^{3}$, conformal Killing spinors satisfying
the equation $\nabla_{\mu}\epsilon=\gamma_{\mu}\eta$ exist, and can be used to
generate the full $\mathfrak{osp}\left(\mathcal{N}|4\right)$ superconformal algebra by embedding $\epsilon_\alpha,\eta_\alpha$ in tensors like $\bar{\Theta}^{IJ}_\alpha$. This algebra includes
diffeomorphisms by Killing, and conformal Killing, vectors as well as Weyl and R-symmetry transformations. In a more general setup, one may generate a supersymmetry using
\emph{generalized Killing spinors} whose equations involve background
fields other than the spin connection and vielbein. Such spinors are
solutions to the equations imposed by the vanishing of the variation
of all fermions in a background supergravity multiplet \cite{Festuccia:2011ws}.

A standard argument shows that an Euclidean path integral with a fermionic symmetry transformation
$Q$, is invariant under $Q$-exact deformations of the
action, as long as these are also $Q$-closed and do not affect
the convergence properties. Convergence requires, in particular, that
$Q^2$ be the generator of a compact bosonic symmetry,
which in turn requires $\mathcal{N}\ge2$ when working on $S^3$. If such deformation
invariance can be used to scale the coupling constants, the calculation
of the integral can be reduced to a calculation in a free 
theory---a process known as localization \cite{Witten:1988ze,Pestun:2007rz}. The result of a localization calculation
is a sum or integral over an often finite dimensional moduli space
of (classical) supersymmetric vacua, with an effective action which
is given exactly by a one loop calculation. 

Supersymmetric actions on $S^{3}$ can be derived by coupling
to supergravity, or by a trial and error process of adding terms
of order $1/r$ and $1/r^{2}$. For a superconformal theory, one only
needs to add the standard conformal mass term for dynamical scalars.
Wilson loops which do not contain fermions can be mapped directly,
since neither the operator nor the variation of the fields appearing
in it contain derivatives. Supersymmetric contours for
such loops are often integral contours of a spinor bi-linear vector
field such as $\bar{\epsilon}\gamma^{\mu}\epsilon$. This is the case
for $W^\text{bos}_{R}$. We now review the evaluation of the exact expectation value of this loop in the $\mathcal{N}=2$ formulation of the ABJ(M) model \cite{Benna:2008zy}.

\subsection{Localization of $\mathcal{N}=2$ gauge theories on the three sphere}

In any $\mathcal{N}=2$ supersymmetric gauge theory on $S^{3}_{r=1}$,
deformation invariance can be used to freely change the value of the
Yang-Mills coupling \cite{Kapustin:2009kz}. The result is localization onto the moduli space
of vector multiplet fields, {\it i.e.}\ the space of solutions to the
equations for the vanishing of the variation of a gaugino
\begin{equation}
\delta\lambda=\left( \frac{1}{2}\gamma^{\mu\nu}F_{\mu\nu}+i \gamma^{\mu}D_{\mu}\sigma-D\right)\epsilon+\sigma\eta.\label{eq:3:gaugino_variation}
\end{equation}
Up to gauge transformations, this moduli space can be parameterized by
an arbitrary spacetime independent profile for the vector multiplet adjoint valued
scalar $\sigma$. The result is therefore a matrix model. No additional moduli arise from chiral multiplets of canonical dimension, due to the conformal mass term. The original action,
and any operator insertions, should be evaluated strictly on this moduli
space.

The one-loop effective action can be derived by expanding all fields to quadratic
order around the moduli space in appropriate
spherical harmonics---eigenfunctions of the various fluctuation operators
$D_{\text{field}}^{\sigma}$---and computing the resulting functional determinant.
Alternatively, one may use the index theorem for transversely elliptic
operators \cite{MR0482866}, which has the advantage of making some cancellations manifest.
In either approach, the result is an infinite product which must be
carefully regularized. We can think of $\sigma$ as defining a Cartan
subalgebra of the gauge Lie algebra. We denote by $\alpha(\sigma)$
the eigenvalue for the action of $\sigma$ on a field proportional
to a root $\alpha$, and by $\rho(\sigma)$ for one proportional
to a weight $\rho$. The fields comprising a vector multiplet, and
proportional to a root pair $\alpha,-\alpha$, including the ghosts
yield an effective action
\begin{equation}
e^{-S_{\text{effective}}\left(\sigma\right)}=\frac{\det D_{\text{gauginos}}^{\sigma}\det D_{\text{ghosts}}^{\sigma}}{\sqrt{\det D_{\text{vector}}^{\sigma}\det D_{\text{scalar}}^{\sigma}}}=\alpha(\sigma)^{2}\prod_{n=1}^{\infty}\left(n^{2}+\alpha(\sigma)^{2}\right)=4\sinh^{2}\pi\alpha(\sigma)\,.
\label{eq:3:vector_multiplet_determinant}
\end{equation}
Similarly, the weight $\rho$ fields in a massive hypermultiplet yield
$\left(2\cosh\pi\left(\rho(\sigma)+m\right)\right)^{-1}$. 

\subsection{Localization in Chern-Simons-matter theories with extended supersymmetry}

The ABJ(M) model is based on the product gauge group $U\left(N_1\right)_k\times U\left(N_2\right)_{-k}$
with hypermultiplets in the $\left(\Box,\bar{\Box}\right)\oplus\left(\bar{\Box},\Box\right)$
representation \cite{Aharony:2008ug,Aharony:2008gk}. Denoting by $\mu_{i}/2\pi,\nu_{i}/2\pi$ the eigenvalues
for the respective $\sigma^{\left(1,2\right)}$ matrices, we have  the following
integral for the expectation value of a 1/6 BPS Wilson loop
in ABJ(M)
\begin{align}
\left\langle W^{\text{bos}}_{R}\right\rangle  & =\underbrace{\frac{1}{N_1!N_2!}}_{\text{normalization}}\underbrace{\int\prod_{i=1}^{N_1}\frac{d\mu_{i}}{2\pi}\prod_{j=1}^{N_2}\frac{d\nu_{j}}{2\pi}}_{\text{moduli space integral}}\underbrace{e^{\frac{i k}{4\pi}\left(\sum_{i}\mu_{i}^{2}-\sum_{j}\nu_{j}^{2}\right)}}_{\text{classical CS action}}\underbrace{\sum_{\rho\in R}e^{\rho\left(\mu,\nu\right)}}_{\text{1/6 BPS Wilson loop}}\label{eq:3:ABJ(M)_matrix_model}\\
 &\quad \underbrace{\prod_{i<j}^{N_1}4\sinh^{2}\frac{1}{2}\left(\mu_{i}-\mu_{j}\right)\prod_{i<j}^{N_2}4\sinh^{2}\frac{1}{2}\left(\nu_{i}-\nu_{j}\right)}_{\text{vector multiplets}}\underbrace{\prod_{i,j}^{N_1,N_2}\left(4\cosh^{2}\frac{1}{2}\left(\mu_{i}-\nu_{j}\right)\right)^{-1}}_{\text{hypermultiplets}}.\nonumber
\end{align}
Above, the contour for $ W^{\text{bos}}_{R}$ is a great circle on $S^3$. Note that $\sigma^{\left(1,2\right)}$ are auxiliary fields in ABJ(M), with {\it e.g.}\ $\sigma^{\left(1\right)}$ equal on-shell
to $\frac{2\pi}{k}{M^I}_{J}C_{I}\bar{C}^J$, which become dynamical in the process
of localization. 

Localization requires off-shell closure of the supercharge. An observable in a model with a larger on-shell realized
supersymmetry algebra, such as ABJ(M), may be annihilated by supercharges
which cannot be rotated into
an $\mathcal{N}=2$ subalgebra. Localization of such an operator requires
closing a single supercharge off-shell, possibly in a non-Poincar{\'e}-invariant
way, by adding appropriate auxiliary fields (see {\it e.g.}\ \cite{Dedushenko:2016jxl}). This approach is currently
being used in order to derive the conjectured matrix model, discussed in Chapter~\ref{chapter:8}, for the
latitude Wilson loop discussed in Chapter~\ref{chapter:7}.

Expectation values of operators within the same cohomology class coincide,
regardless of off-shell closure. The 1/2 and 1/6 BPS fermionic Wilson loops of Chapter~\ref{chapter:2}
can be shown to be cohomologically
equivalent to the bosonic 1/6 BPS Wilson loop within the ABJ(M) model \cite{Drukker:2009hy}, {\it i.e.}\ there
exists a well defined operator $V$, a supercharge $Q$, and a map $R\rightarrow \mathcal{R}$ such that 
\begin{equation}
 W^\text {fer}_{\mathcal{R}}= W^\text {bos}_{R}+Q V.
\end{equation}
A similar statement holds for the versions of
the latitude loop reviewed in Chapter~\ref{chapter:7}. 

\subsection{Acknowledgments}

The work presented here was carried out in collaboration with A.~Kapustin and B.~Willett. 
The work of IY is supported by Italian
Ministero dell\textquoteright Istruzione, Universit\`a e Ricerca (MIUR),
and Istituto Nazionale di Fisica Nucleare (INFN) through the Gauge
and String Theory (GAST) research project.

\ifaffiliation
\subsection{Full affiliation}
Itamar Yaakov, Dipartimento SMFI, Universit\`a di Parma and INFN Gruppo
Collegato di Parma. Parco Area delle Scienze, 7/A, 43124 Parma PR,
Italy.
\fi
\authors{Pavel Putrov\ifaffiliation, ICTP\fi}
\section{Solution of the ABJM matrix model}
\label{chapter:4}
\subsection{Background}
In this chapter we take a close look at the matrix model that arises from localization of the ABJM theory with possible 1/6 or 1/2 BPS Wilson loop insertion, as reviewed in Chapter~\ref{chapter:3}. There are two natural limits when the matrix model can be analyzed
\begin{itemize}
    \item the 't~Hooft limit $N_1,N_2 \rightarrow \infty$, and $k\rightarrow 
\infty$ while the 't~Hooft parameters $\lambda_i=N_i/k$ are kept fixed. This limit is natural for comparison with the holographic dual type IIA string theory on $AdS_4\times \mathbb{CP}^3$. The string coupling constant is related to the level as $g_s=2\pi i/k$.
 \item the ``M-theory limit'' $N_1,N_2 \rightarrow \infty$ while the level $k$ is kept fixed. This limit is natural for comparison with the holographic dual M-theory on $AdS_4\times S^7$, where the radius of the circle fiber of the Hopf fibration $S^7\rightarrow \mathbb{CP}^3$ is $1/k$.
\end{itemize}

\subsection{'t~Hooft limit solution}
In the 't~Hooft limit the matrix model can be analyzed by the rather standard and general technique of spectral curve and topological recursion \cite{Marino:2009jd,Drukker:2010nc}. The same matrix model, up to an analytic continuation in ranks $N_i$, was previously considered in the context of bosonic Chern-Simons on a Lens space and topological strings on local $\mathbb{P}^1\times \mathbb{P}^1$ \cite{Aganagic:2002wv,Halmagyi:2003ze,Halmagyi:2003mm}. The main idea is that in the 't~Hooft limit the leading (also known as \textit{planar}) contribution is given by the configurations where eigenvalues $e^{\mu_i}$ and $e^{\nu_i}$ in the matrix integral are distributed along two intervals in the complex planes. These intervals can be found by solving the equations of motion for an eigenvalue in the background of all the other eigenvalues and can be interpreted as cuts of a Riemann surface spread over the complex plane. The Riemann surface is described by an algebraic equation
\begin{equation}
    Y+X^2Y^{-1}-\beta(X^2+i\kappa X-1)=0,\qquad X,Y\in \mathbb{C}^*,
    \label{4:spectral-curve}
\end{equation}
and is referred to as a \textit{spectral curve}. The density of the eigenvalues along the cuts is given by the jumps of a meromorphic 1-form across the cut. Namely, if one defines the \textit{resolvent} 1-form as the following expectation value in the matrix model
\begin{equation}
    \omega:=g_s\left\langle 
    \sum_{i=1}^{N_1}\frac{X+e^{\mu_i}}{X-e^{\mu_i}}
    -\sum_{i=1}^{N_2}\frac{X-e^{\nu_i}}{X+e^{\nu_i}}
    \right\rangle_\text{MM} \frac{dX}{X},
\end{equation}
it admits a genus expansion of the following form
\begin{equation}
    \omega=\sum_{g\geq 0}g_s^{2g}\omega_g,
    \qquad
    \omega_0=\log Y \,\frac{dX}{X}.
\end{equation}
\begin{figure}[t]
\centering
\includegraphics[width=.5\textwidth]{\folder/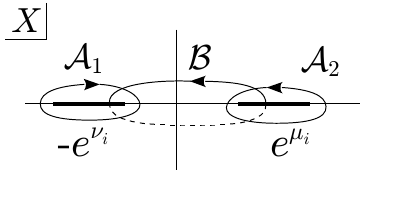}
\caption{The distribution of the eigenvalues in the ABJM matrix model and the contours in the $X$-plane.}
\label{4:fig-contours}
\end{figure}
The complex parameters $\beta$ and $\kappa$ of the curve are fixed by the two equations $\lambda_i=\int_{\mathcal{A}_i}\omega_0$, where the contours $\mathcal{A}_i$ encircle the cuts (see Figure~\ref{4:fig-contours}). Any one-point correlation function in the matrix model can be expressed via the resolvent. For example, the expectation value of 1/6 BPS Wilson loop in the fundamental representation of $U(N_1)$ is given by ({\it cf.}\ \eqref{2:bosonic}, (\ref{eq:3:ABJ(M)_matrix_model}))
\begin{equation}
   \left\langle W_{(\square,\mathbf{1})}^\text{bos} \right\rangle = \Big\langle \sum_i e^{\mu_i}\Big\rangle_\text{MM}
    =\frac{1}{2\pi ig_s}\int_{\mathcal{A}_1}X\,\omega 
    = \sum_{g\geq 0}g_s^{2g-1}\left\langle W_{(\square,\mathbf{1})}^\text{bos} \right\rangle_{g},
    \label{4:W-integral}
\end{equation}
where the last equality is its genus expansion. From now on, for the technical simplicity of the formulas let us assume that $N_1=N_2=N$  so that there is a single 't~Hooft parameter $\lambda_1=\lambda_2=\lambda$.  For more general expressions we refer to the papers cited. It follows that $\beta=1$ in the spectral curve equation (\ref{4:spectral-curve}). The relation between the 't~Hooft parameter and the parameter $\kappa$ of the elliptic curve then explicitly reads
\begin{equation}
\label{4:kappa}
 \lambda=\frac{\kappa}{8 \pi}   {~}_3F_2\left(\frac{1}{2},\frac{1}{2},\frac{1}{2};1,\frac{3}{2};-\frac{\kappa^2
   }{16}\right).
\end{equation}
The planar limit of the expectation value of the 1/6 BPS Wilson loop can be determined by the following equation
\begin{equation}
    \frac{\partial\big\langle W_{(\square,\mathbf{1})}^\text{bos}\big \rangle_0  }{\partial \kappa}=
    -\frac{i}{\pi\sqrt{ab}(1+ab)}(aK(s)-(a+b)\Pi(n|s)),
\end{equation}
where $K(s)$ and $\Pi(n|s)$ are complete elliptic integrals of the first and third kind respectively, and $a,b,s$ and $n$ are the related to the curve parameter $\kappa$ as follows
\begin{equation}
    a= \frac{1}{2}\left(2+i\kappa+\sqrt{\kappa(4i-\kappa)}\right),\qquad
    b= \frac{1}{2}\left(2-i\kappa+\sqrt{-\kappa(4i+\kappa)}\right), 
    \label{4:ab-kappa}
\end{equation}
\begin{equation}
    s^2=1-\left(\frac{a+b}{1+ab}\right)^2,\qquad n=\frac{b}{a}\,\frac{a^2-1}{1+ab}.
\end{equation}
The free energy $F=\log Z$ also admits the genus expansion $F=\sum_{g\geq 0}g_s^{2g-2}F_g(\lambda)$
with the planar limit determined by another period of the resolvent
\begin{equation}
    \frac{\partial F_0(\lambda)}{\partial \lambda} = \int_{\mathcal{B}}\omega_0 =
    2\pi^2 \log \kappa +\frac{4 \pi^2}{\kappa^2} \, {}_4 F_3 \left( 1, 1, \frac{3}{2}, \frac{3}{2}; 2,2,2; -\frac{16}{\kappa^2}\right),
\end{equation}
where the contour $\mathcal{B}$ is shown in Figure~\ref{4:fig-contours}. The strong coupling limits of the planar contributions to the free energy and the Wilson loop take the following form
\begin{equation}
    \partial\left\langle W_{(\square,\mathbf{1})}^\text{bos} \right\rangle_0 \sim e^{\pi\sqrt{\lambda}},\qquad
    F_0(\lambda)\sim \frac{4\pi^3\sqrt{2}\lambda^{3/2}}{3},\qquad \lambda\rightarrow\infty,
\end{equation}
which is in perfect agreement with AdS$_4$/CFT$_3$ correspondence. The strong coupling limit can be also obtained directly by the methods of \cite{Herzog:2010hf} or \cite{Santamaria:2010dm}. The higher genus corrections can be produced order-by-order via topological recursion procedure of \cite{Eynard:2007kz} or by solving holomorphic anomaly equations \cite{Drukker:2010nc,Drukker:2011zy}.

\subsection{M-theory solution via Fermi gas representation}

In \cite{Marino:2011eh} a different approach to solving the ABJM matrix model was developed. It is well suited to study the partition function and the Wilson loop directly in the ``M-theory limit''. The main idea is presentation of the matrix model as a partition function of a free Fermi gas with a certain 1-particle Hamiltonian. In particular this leads to direct derivation of the exact perturbative (w.r.t.  $1/N$) contribution to the partition function
\begin{equation}
 Z(N)=e^{A}\text{Ai}\left[C^{-1/3}\left(N-B\right)\right] \left(1+{\cal O}(e^{-c\sqrt{N}})\right),
 \label{4:partitionfunction}
\end{equation}
where $\text{Ai}$ is the Airy function of the first kind, $C=2/\pi^2 k$, $B=k/24+1/3k$, and $A$ is a certain special function of $k$. The same result was first obtained by resummation of the genus expansion in the 't~Hooft limit in \cite{Fuji:2011km}.

\subsection{Acknowledgements}
The author is thankful to N.~Drukker and M.~Mari\~no for collaboration on the topic.

\ifaffiliation
\subsection{Full affiliation}
ICTP, Strada Costiera 11, Trieste 34151, Italy.
\fi
\authors{Marco S. Bianchi\ifaffiliation, Universidad Austral de Chile\fi}
\section{Perturbative methods for the 1/2 BPS Wilson loop}\label{chapter:5}

\subsection{State of the art}

Two loops is the state of the art for the perturbative determination of the 1/2 BPS Wilson loop expectation value.
The relevant one- and two-loop diagrams, based on the definition of the operator of \cite{Drukker:2009hy}, evaluate \cite{Bianchi:2013zda,Bianchi:2013rma,Griguolo:2013sma,Drukker:2008zx,Chen:2008bp,Rey:2008bh}
\begin{equation}
\begin{gathered}
\label{5:break}
\raisebox{-0.55cm}{\includegraphics[width=1.5cm]{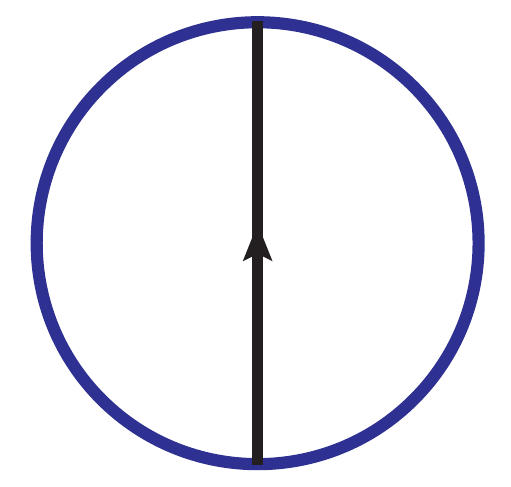}} = \raisebox{-0.55cm}{\includegraphics[width=1.5cm]{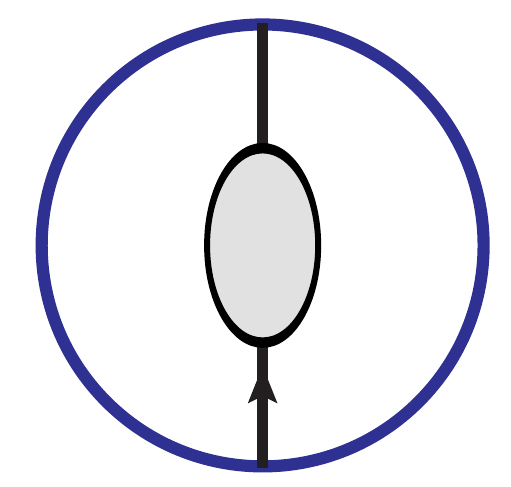}} = 0, 
\qquad 
\raisebox{-0.55cm}{\includegraphics[width=1.5cm]{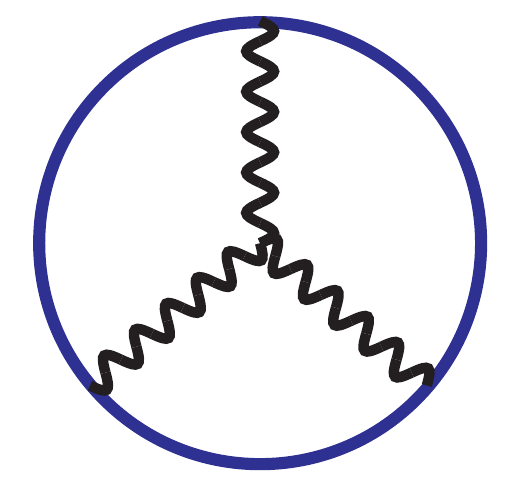}} = \displaystyle
-\frac{\pi^2}{6}\,\frac{N_1+N_2}{k^2} \left(N_1^2+N_2^2-N_1N_2-1\right),
\\
\raisebox{-0.55cm}{\includegraphics[width=1.5cm]{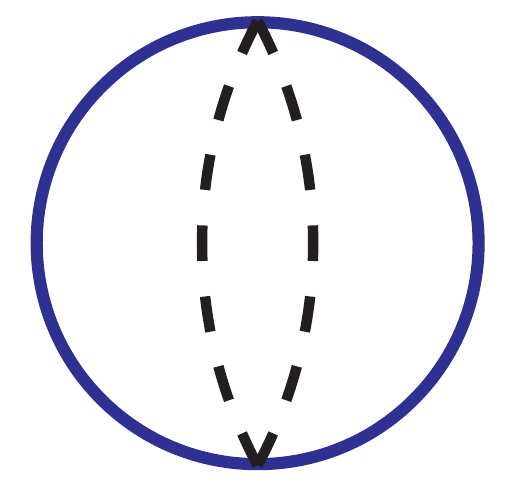}} + \raisebox{-0.55cm}{\includegraphics[width=1.5cm]{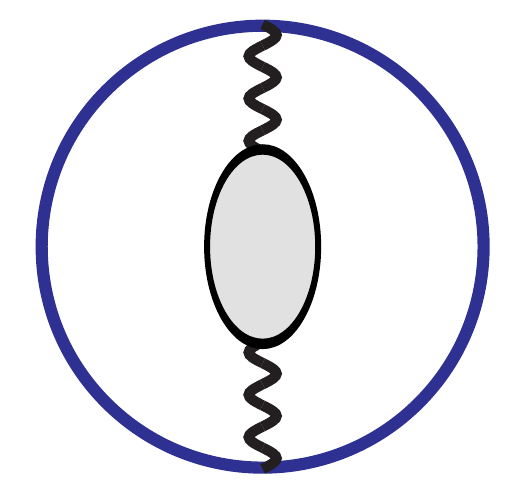}} = \displaystyle\frac23\, \raisebox{-0.55cm}{\includegraphics[width=1.5cm]{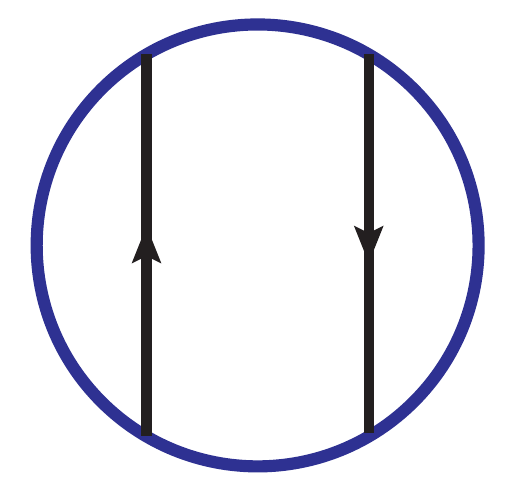}} = -\frac12 \raisebox{-0.55cm}{\includegraphics[width=1.5cm]{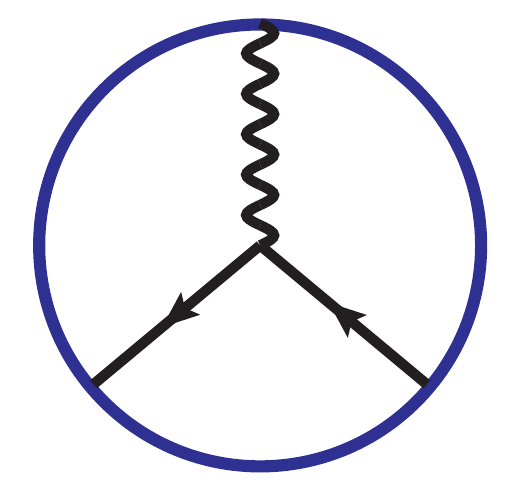}} = \displaystyle \pi^2 N_1N_2 \,\frac{N_1+N_2}{k^2}.
\end{gathered}
\end{equation}
Solid, dashed and wavy lines stand for fermions, scalars and gluons, respectively and blobs represent one-loop self-energy corrections.
The final two-loop result for the expectation value reads
\begin{equation}\label{5:pert}
\left\langle W_{1/2} \right\rangle = \left(N_1 + N_2\right) \left[ 1 - \frac{\pi^2}{6 k^2} \left( N_1^2 + N_2^2 - 4N_1 N_2 -1 \right) + {\cal O}\left(k^{-4}\right) \right].
\end{equation}
It coincides with the prediction from localization, after removing a framing phase $e^{{i \pi}( N_1 - N_2)/k}$ from the latter (see Chapter~\ref{chapter:6} on this issue).
This agreement provides a successful test of localization in ABJM.

The computation outlined above can be extended to include multiple windings \cite{Bianchi:2016gpg} and, equivalently, different gauge group representations. 
Its backbone can be utilized for evaluating the two-loop expectation value of similar fermionic operators, existing in  quiver Chern-Simons-matter theories with lower supersymmetry of Chapter~\ref{chapter:9}.
Moreover, the same diagrams appear, albeit with some deformations, when determining the expectation value of latitude Wilson loops in ABJM, discussed in Chapters~\ref{chapter:7}-\ref{chapter:8}.

\subsection{New versus old formulation}

The reformulation of the 1/2 BPS operator presented in Chapter~\ref{chapter:2} differs from \cite{Drukker:2009hy} in two aspects: additional $\pm{|\dot{x}|}/{4|x|}$ terms in the diagonal entries of the superconnection \eqref{2:superconnection} and different fermionic couplings $\eta$, $\bar\eta$. On the maximal circle ${\cal C}: \{ 0, \cos\tau, \sin\tau \}$, $0\leq \tau<2\pi$, the former evaluate to $\pm1/4$ and the spinors read
\begin{equation}
\eta^\alpha_I = \frac{1}{\sqrt{2}} \left(
1\quad  {-i} e^{-i \tau}
\right) \delta_I^1, \qquad
\bar{\eta}^I_\alpha = \frac{1}{\sqrt{2}}
\begin{pmatrix}
1\\
i e^{i \tau}\\
\end{pmatrix} \delta^I_1.
\end{equation}
These modifications produce a different perturbative expansion, which eventually re-sums into that of the old formulation.
To see this explicitly, let us consider the fermion exchange diagram at one-loop.
The following shorthand notation is used
$x_i\equiv x(\tau_i)$, $x_{ij}\equiv x_i-x_j$ and $\eta_i\equiv \eta(\tau_i)$.
Expanding the exponential of \eqref{2:loop} to second order gives the fermion contractions
\begin{align}\label{5:eq1}
& -4\pi \frac{N_1N_2}{k}
\int_0^{2\pi}d\tau_1\int_0^{\tau_1} d\tau_2
 \left\langle (\eta\bar\psi)_1 (\psi\bar\eta)_2 - (\psi\bar\eta)_1(\eta\bar\psi)_2 \right\rangle =
i 2^{2-2\epsilon} K
 \frac{\left(\eta_1 \gamma_{\mu}  \bar{\eta}_2 - \eta_2 \gamma_{\mu}  \bar{\eta}_1 \right) x_{12}^{\mu}}{(x_{12}^2)^{3/2-\epsilon}},
\end{align}
where dimensional regularization ($d=3-2\epsilon$) is used in the fermion propagator and an ubiquitous factor is defined $K\equiv \frac{\Gamma\left(3/2-\epsilon\right)}{2^{1-2\epsilon}\pi^{1/2-\epsilon}} \frac{N_1N_2}{k}$.
An integral of trigonometric functions is obtained, that can be performed turning them into infinite sums, according to the method described in \cite{Bianchi:2013rma}
\begin{equation}
\label{5:eq2}
\int_0^{2\pi}d\tau_1\int_0^{\tau_1} d\tau_2 \frac{1-\cos(\tau_1-\tau_2)}{\left(\sin^2\frac{\tau_1-\tau_2}{2}\right)^{3/2-\epsilon}} = 
\frac{4 \pi ^{3/2} \Gamma (\epsilon )}{\Gamma \left(1/2+\epsilon\right)}.
\end{equation}
This result exhibits a $1/\epsilon$ pole and differs from the analogous evaluation in the original papers, using the old formulation. The discrepancy lies in the fact that, at fixed $1/k$ order, infinite additional contributions arise expanding the exponential \eqref{2:loop} in the new definition. They come from insertions of the constant terms $\pm 1/4$ in the diagonal entries of the superconnection. They do not depend on the coupling, hence any number of them contributes to the same perturbative order. The perturbative series is thus re-organized in terms of a new expansion in the number of $\pm1/4$ insertions. For the fermion exchange diagram it reads explicitly
\begin{align}
\raisebox{-0.55cm}{\includegraphics[width=1.5cm]{ferm1}} &= \sum_{l=0}^\infty \frac{A_l}{4^l} \,,
\\
A_l 
&=\frac{4\pi}{k}N_1N_2
\int_{0}^{2\pi}d\tau_1 \dots \int_{0}^{\tau_{l+1}}d\tau_{l+2}
 \sum_{j>i} (-1)^{j-i} \sTr\begin{pmatrix}
 (-i)^l \left\langle(\eta\bar\psi)_i (\psi\bar\eta)_j\right\rangle & \dots\\
 \dots & i^l\left\langle(\psi\bar\eta)_i (\eta\bar\psi)_j\right\rangle \end{pmatrix}.
\nonumber
\end{align}
Then, \eqref{5:eq2} is the zeroth order in this alternative expansion, $A_0$. Let us separate terms from the upper-left and lower-right  blocks, before taking the supertrace, $A_l = A_l^{(+)} - A_l^{(-)}$.
A general expression for 
$A_l^{(+)} = -\big(A_l^{(-)}\big)^*$ can be derived in terms of generalized zeta functions  $\zeta_s(x)\equiv \sum_{k=0}^\infty (k+x)^{-s}$ 
\begin{align}
& A_l^{(+)} = \frac{K (-i)^{l} \pi ^l}{2^{-l-1}l!} \begin{cases}
 \frac{\pi -2 i (l+1)}{\epsilon}+ 2\pi  \log 2+i (\pi  (\pi +2 i) l-4 (l+1) (1+\log 2)) & \text{$l$ even} \\
\frac{\pi -2 i l}{\epsilon} + 2 \pi  l^2+ i l \left(\pi ^2-4 (1+\log 2)\right)+\pi  (2\log 2 -2-i \pi)-\frac{14 l(l+1) \zeta_3}{\pi }  & \text{$l$ odd}
\end{cases}
\nonumber\\&~~
-K \sum _{m=0}^{\lfloor\frac{l-2}{2}\rfloor} \frac{(-1)^m \pi ^{2 m} \left(2 \pi \zeta_{l-2 m}(\frac{3}{2})+(2 i (l+1)-\pi) \zeta_{l-2 m+1}(\frac{3}{2})-i (l+1) \zeta_{l-2 m+2}(\frac{3}{2})\right)}{2^{-l-2}(2 m)!}
+ {\cal O}(\epsilon).
\end{align}
To retrieve the complete result, such contributions can be re-summed efficiently at the level of the integrand. Performing combinatorics and contour integral manipulations, the $\pm1/4$ expansion is cast into the form
\begin{equation}\label{5:eq3}
\raisebox{-0.55cm}{\includegraphics[width=1.5cm]{ferm1}}=  
\frac{K}{2}
\int_0^{2\pi}d\tau_1\int_0^{\tau_1} d\tau_2  
\left[ \sum_{l=0}^{\infty} \frac{(-i)^l}{l!}\left(\frac{\pi - \tau_1 +\tau_2}{2} \right)^l  \right]
\frac{1-e^{-i (\tau_1-\tau_2)}}{\left(\sin^2\frac{\tau_1-\tau_2}{2}\right)^{3/2-\epsilon}} + \text{c.c.}.
\end{equation}
Summing over $l$
reproduces the factor appearing in the fermion exchange diagram in the old formulation \cite{Bianchi:2013rma}
\begin{equation}
\raisebox{-0.55cm}{\includegraphics[width=1.5cm]{ferm1}} =
2K
\int_0^{2\pi}d\tau_1\int_0^{\tau_1} d\tau_2  
\frac{1}{\left(\sin^2\frac{\tau_1-\tau_2}{2}\right)^{1-\epsilon}} = 
-2^{1+2\epsilon}\pi ^{2+\epsilon} \frac{N_1N_2}{k}\frac{ \sec (\pi  \epsilon )}{\Gamma (\epsilon )} = {\cal O}(\epsilon).
\end{equation}
The sign swaps across the upper-left and lower-right 
blocks ultimately reconstruct the twist matrix of \cite{Drukker:2009hy}.

In conclusion, the perturbative expansion of fermionic graphs in the old and new formulations differ substantially, but they prove to be equivalent. This should be the case, being they related by a gauge transformation.
The new definition appears more natural and elegant in some respects, spelled out in Chapter~\ref{chapter:2}.  However its perturbative expansion entails re-summations as in \eqref{5:eq3}, which in the original definition of \cite{Drukker:2009hy} are already built-in. The latter is thus more compact and better suited for the purpose of a perturbative evaluation.

\subsection{Future directions}

The two-loop computation of fermionic diagrams has only been performed at trivial framing.
Conversely, the cohomological equivalence described in Chapter~\ref{chapter:3} imposes constraints on the expectation values at framing one. 
These imply that fermionic diagrams must possess non-trivial framing contributions, however a direct computation thereof is lacking.
It would be interesting to improve the efficiency of the perturbative expansion by adopting a superspace description, which has not been developed thus far.
Finally, a more direct test of localization would entail directly considering the theory on a curved background, but perturbation theory in this setting seems challenging with current technology.

\ifaffiliation
\subsection{Full affiliation}
Universidad Austral de Chile, Casilla 567, Valdivia, Chile
\fi

\authors{Matias Leoni\ifaffiliation, Universidad de Buenos Aires \& IFIBA--CONICET\fi}
\section{Framing in Chern-Simons-matter theories}
\label{chapter:6}

\subsection{Pure Chern-Simons}

In its simplest formulation, pure Chern-Simons theory can be written down without reference to any metric of the manifold where it is formulated. One could naively think that this would mean that any gauge invariant observable defined for this theory should depend only on topological properties of the manifold, the gauge group, the Chern-Simons level and the observable in question. This expectation turns out to be too optimistic: It is a well-known fact \cite{Witten:1988hf,Guadagnini:1989am,Alvarez:1991sx} that when a regularization of the theory is introduced, some extra data is needed. In particular, for ordinary Wilson loops defined in pure Chern-Simons theory this extra information is the \textit{framing} of the paths in the loop.

Consider the basic Wilson loop in pure Chern-Simons theory defined on a closed path $\Gamma$ in the fundamental representation of $U(N)$: $\langle W_{\Gamma} \rangle=\left\langle \Tr \mathcal{P} \exp \left(i\oint_\Gamma A_\mu dx^\mu \right) \right\rangle$.
The only Feynman diagram at leading order in perturbation theory is a simple contractible gauge propagator $\langle A_\mu(x_1)A_\nu (x_2)\rangle$ joining two points of the curve with both points integrated over the curve. To avoid divergences when both points collide we choose to deform the path of the second gauge field (which we call $\Gamma_f$) in the following way
\begin{equation}
x^\mu(\tau)\rightarrow y^\mu(\tau)=x^{\mu}(\tau)+\delta\, n^{\mu}(\tau),
\end{equation}
with $n^{\mu}(\tau)$ a unit vector and $\delta$ a parameter which we may eventually take to be arbitrarily small. In a sense we can think that instead of a curve travelled by both transported fields, we are defining a two-dimensional object, a ribbon. 
When removing the regularization by taking $\delta\to 0$, we are left with an ambiguity. 
The straightforward computation leads to the following integral
\begin{equation}
\text{link}(\Gamma,\Gamma_f)=\frac{1}{4\pi}\oint_\Gamma dx^\mu \oint_{\Gamma_f} dy^\nu \epsilon_{\mu\nu\rho}\frac{(x-y)^{\rho}}{|x-y|^3}=f.
\end{equation}
This is the very well-known Gauss linking integral topological invariant. This means that even if we took the limit of $\delta\to 0$ when both curves coincide, the twisting of one curve over the other survives the limit and the ambiguity is just the number of times ($f$) one of the curves winds over the other. This is the extra piece of information one has to add to the definition of the Wilson loop (see \cite{Moore:TASI} for a thorough review).

While we just explained the leading appearance of framing, it was shown in \cite{Alvarez:1991sx} that framing-dependent parts of the Wilson loop computation exponentiate in a controlled way in pure Chern-Simons theory
\begin{equation}\label{6:FramingExpo}
\langle W_{\Gamma}\rangle_f=e^{\frac{i\pi N}{k}\text{link}(\Gamma,\Gamma_f)}
\langle W_{\Gamma}\rangle_{f=0}.
\end{equation} 

\subsection{Enter matter}
While the framing subtlety is clearly important to understand to which degree pure Chern-Simons theory is a topological theory, when matter is coupled to it and the theory becomes an ordinary non-topological theory as  ABJ(M), one could in principle choose to put framing to zero. The main reason why framing is still relevant in Chern-Simons-matter theories is because the localization technique for finding exact results of Wilson loops returns results with non-vanishing framing. Specifically, the result from localization, as presented in Chapters~\ref{chapter:3} and~\ref{chapter:4} is necessarily at framing 1 for both 1/2 BPS \cite{Drukker:2009hy} and 1/6 BPS  \cite{Drukker:2008zx,Chen:2008bp,Rey:2008bh} Wilson loops. This is because the regularization compatible with supersymmetry, when one localizes the function integral on $S^3$, has the path and its frame wrapping different Hopf fibers\cite{Kapustin:2009kz}.
 
The authors of \cite{Drukker:2009hy} showed that the 1/2 BPS Wilson loops and a combinations of the bosonic 1/6 BPS pairs are in the same cohomology class under the localizing supercharge. In \cite{Bianchi:2013zda,Bianchi:2013rma} evidence was given that this cohomological equivalence is realized at the quantum level specifically at framing $1$, once again emphasizing the importance of framing for non-topological realizations of Chern-Simons theories. The two-loop results of \cite{Bianchi:2013zda,Bianchi:2013rma} also show that at least to that perturbative order, the relation between framing $1$ and framing $0$ quantities is a simple phase necessary for consistency with the cohomological equivalence.

The authors of \cite{Bianchi:2016yzj} went further and made a combined analytical/numerical analysis of the 1/6 BPS Wilson loop \eqref{2:bosonic} up to third order in perturbation theory. The result for arbitrary framing $f$ is consistent with the expansion
\begin{equation}\label{6:1sixthBPS}
\left\langle W^\text{bos} \right\rangle_f=
e^{i\pi\left(\lambda_1-{\pi^2}\lambda_1\lambda_2^2/2+\mathcal{O}(\lambda^5)\right)f}
\left(1-\frac{\pi^2}{6}\left(\lambda_1^2-6\lambda_1\lambda_2\right)+\mathcal{O}(\lambda^4)\right),
\end{equation}
where $\lambda_1=N_1/k$, $\lambda_2=N_2/k$ are the 't~Hooft couplings of ABJ theory. This result deserves some comments: Firstly, while the framing-dependent contributions seem to exponentiate as in \eqref{6:FramingExpo}, the exponent becomes a non trivial function of the coupling, as opposed to the simple linear exponent of pure Chern-Simons theory; secondly, the analysis of \cite{Bianchi:2016yzj} shows that while up to two-loops all the framing contributions came from purely gauge contractible propagators, at three-loops vertex-like diagrams with matter also contribute to the framing anomaly. An interesting consequence of the non-triviality of the exponent of \eqref{6:1sixthBPS} has to do with the fact \cite{Bianchi:2014laa,Bianchi:2017svd,Bianchi:2018bke,Bianchi:2017ozk} that the Bremsstrahlung function (Chapter~\ref{chapter:10} and \ref{chapter:11}) associated to 1/2 BPS Wilson loops in ABJM theory ($N_1=N_2$) can be written as $B_{1/2}=(8\pi)^{-1}\tan\Phi_B$ where $\Phi_B$ is the complex phase of the 1/6 BPS Wilson loop at framing $1$: this implies a curious and very intimate connection between framing and the Bremsstrahlung physics of the theory which deserves further study.

Finally, framing also plays an important, albeit odd role in the DGRT-like construction \cite{Cardinali:2012ru} of Wilson loops in ABJ(M) theory (see Chapter~\ref{chapter:7}). The latitude deformations of both the 1/2 BPS and bosonic 1/6 BPS Wilson loops belong to the same cohomology class \cite{Bianchi:2014laa}. More specifically, it was shown in \cite{Bianchi:2014laa} from a perturbative computation that in order to realize the cohomological equivalence at the quantum level, the analysis has to be done at arbitrary framing $f$ and then a formal identification of the integer $f$ with the effective latitude parameter $q=\sin 2\alpha\cos\theta_0$ has to be performed (here $\alpha$ and $\theta_0$ are angles that characterize the geometry of the latitude Wilson loops). This is an awkward choice since $f$ is an integer number whereas $q$ is a real one.  This is however supported by the matrix model construction of \cite{Bianchi:2018bke}, where a single $q$ parameter is needed in the matrix integral and operator definitions in order to match the known perturbative results with $f=q$ (see Chapter~\ref{chapter:8}).

\subsection{Acknowledgements}
We would like to thank M.S. Bianchi, G. Giribet, L. Griguolo, A. Mauri, S. Penati and D. Seminara for collaborations on the subjects presented here.

\ifaffiliation
\subsection{Full affiliation}
Departamento de F\'isica, Universidad de Buenos Aires \& IFIBA - CONICET. Ciudad Universitaria, pabell\'on 1 (1428) Buenos Aires, Argentina.
\fi

\authors{Luca Griguolo\ifaffiliation, Universit\`a di Parma and INFN Parma\fi}
\section{BPS Wilson loops with more general contours}
\label{chapter:7}

\subsection{Background}
Four-dimensional ${\cal N}=4$ SYM theory admits a variety of different BPS Wilson loops, that generalize non-trivially the original straight line and circular ones \cite{Erickson:2000af,Drukker:2000rr}. A complete classification of admissible contours with the appropriate scalar couplings has been performed in \cite{Dymarsky:2009si}, exploiting the properties of the relevant Killing spinors. Two important classes of loops have been widely studied and used to derive interesting results: the so-called Zarembo loops \cite{Zarembo:2002an} and the DGRT loops \cite{Drukker:2007qr}. A subset of DGRT operators, preserving 1/8 of the original supersymmetry, are contained in an $S^2$ and their quantum behavior is governed by perturbative 2d Yang-Mills theory \cite{Drukker:2007qr,Pestun:2009nn,Bassetto:1998sr}.

In this chapter two families of BPS fermionic Wilson loops in ABJM theory are described: They can be considered the analogs of the Zarembo and DGRT loops in three dimensions, their bosonic and fermionic couplings depending non-trivially on their path \cite{Cardinali:2012ru}. The key idea, already exploited to construct 1/6 BPS fermionic circles in Chapter~\ref{chapter:2}, is to embed the natural $U(N_{1})\times U(N_{2})$ gauge connection present in ABJM theory into a superconnection parameterized by the path-dependent functions $M_{J}^{ I}$, $\eta_{I}^{\alpha}$ and $\bar{\eta}^{I}_{\alpha}$ (see \eqref{2:superconnection}). The strategy is to derive first a general set of algebraic and differential conditions for them that guarantee the local preservation of a fraction of supersymmetry, up to total derivative terms along the contour. Then one imposes that solutions of these constraints can be combined into a conformal Killing spinor $\bar\Theta^{IJ}=\bar\theta^{IJ}-(x\cdot\gamma)\bar\epsilon^{IJ}$, where $\bar\theta^{IJ}$ and $\bar\epsilon^{IJ}$ are constant. Finally the total derivative terms, organizing into a supergauge transformation, should become irrelevant by taking the super-trace of the Wilson loop operator. This last step requires in general to improve the bosonic part of the connection with a background term, as done in Chapter~\ref{chapter:2} (see \eqref{2:superconnection}), curing the non-periodicity of the couplings and avoiding the presence of the twist-matrix originally introduced in \cite{Cardinali:2012ru}. 

\subsection{Zarembo-like Wilson loops}
This is a family of Wilson loops of arbitrary shape, which preserve at least  one supercharge  of Poincar\'e type, {\it i.e.}\ a supercharge with $\bar\epsilon^{IJ}=0$. These
operators can be viewed as the three-dimensional companion of the loops discussed in \cite{Zarembo:2002an} and a generalization of the BPS straight-line constructed in  \cite{Drukker:2009hy}, which is the simplest example enjoying this property.  In this case, the differential condition is solved trivially and the problem is completely fixed by choosing four constant spinors $s^{I}_{\alpha}$ with completeness relation $\bar s^{I}_{\beta} s_{I}^{\alpha}=\delta^{\alpha}_{\beta}$. Defining $\Pi_+$ as in \eqref{2:eta}, the general form of the couplings is obtained \cite{Cardinali:2012ru}
\be
 \eta_I = s_I \Pi_+, \qquad
  \bar{\eta}^I =
  \Pi_+ \bar{s}^I, \qquad
  M_I^J=\delta_I^J- {s}_I \bar{s}^J-\frac{\dot{x}_\mu}{|\dot{x}|}{s}_I \gamma^\mu\bar{s}^J. 
  \ee
The loops are generically 1/12 BPS and the finite supergauge transformations generated by the relevant supersymmetry transformations are well-defined on any closed contour in $\mathbb{R}^3$. Taking the super-trace (see \eqref{2:loop}) a SUSY-invariant operator is obtained without introducing background terms. The explicit form of the superconnection is
\begin{gather}
\label{2:superconn1}
\mathcal{L}= \begin{pmatrix}
  \mathcal{A}^{(1)} &&
  \sqrt{-\frac{4\pi i }{k}}|\dot x| \eta_I\bar{\psi}^I\\
  \sqrt{-\frac{4\pi i}{k}} |\dot x|\psi_I\bar{\eta}^I &&
  \mathcal{A}^{(2)}\\
  \end{pmatrix},\qquad 
  \begin{matrix}
  \mathcal{A}^{(1)} = \cA^{(1)}_\mu\dot{x}^\mu
  - \frac{2\pi i}{k} |\dot{x}| M_J^I C_I \bar{C}^J \,,\\[3pt]
  \mathcal{A}^{(2)} = \cA^{(2)}_\mu\dot{x}^\mu
  - \frac{2\pi i}{k} |\dot{x}| M_J^I \bar{C}^J C_I.
  \end{matrix}
\end{gather}

\subsection{DGRT-like Wilson loops}
A second family of Wilson loops, defined for an arbitrary curve on the unit sphere $S^{2}\subset\bR^3$ ($x^\mu x_\mu=1)$, can be easily obtained from the previous one.  The central idea is to introduce a matrix $U$ constructed with the coordinates $x^{\mu}(\tau)$ of the circuit, namely $U=\cos\alpha+i(x^{\mu}\gamma_{\mu})\sin\alpha$, with $\alpha$ a free constant angular parameter. Defining an auxiliary $constant$ supercharge $\bar\Delta^{IJ}=\bar\Theta^{IJ}U$ and introducing a background term in the bosonic part of the connection as in Chapter~\ref{chapter:2}
\begin{equation}
\begin{array}{c}
\mathcal{A}^{(1)} = A^{(1)}_{\mu} \dot{x}^\mu
- \frac{2\pi i}{k} |\dot{x}| M_J^I C_I \bar{C}^J
+ \frac{q |\dot{x}|}{4},\\[3pt]
\mathcal{A}^{(2)} = A^{(2)}_{ \mu} \dot{x}^\mu
- \frac{2\pi i}{k} |\dot{x}| M_J^I \bar{C}^J C_I
-\frac{q |\dot{x}|}{4},
\end{array} \qquad
q \equiv \sin{2\alpha},
\label{superconn2}
\end{equation}
the problem becomes formally equivalent to the Zarembo-like one \cite{Cardinali:2012ru}. The solution for the couplings, preserving the superconformal charge $\bar\Theta^{IJ}$, is obtained by rotating the spinors $s^{I}_{\alpha}$ appearing in the Zarembo-like solution
\begin{equation}
\label{DGRTc}
\eta_I = U^\dagger{s_I} \Pi_+, \qquad
\bar{\eta}^I = \Pi_+ \bar{s}^I U, \qquad
M^I_J =\delta^I_J -
s_J \left( 1 + \frac{\dot{x}^\mu \gamma_\mu}{|\dot{x}|}\cos2\alpha +
x^\mu \frac{\dot{x}^\nu}{|\dot{x}|} \gamma^\rho \epsilon_{\mu\nu\rho} \sin2\alpha \right)
\bar{s}^I.
\end{equation}
A particularly interesting example of DGRT-like Wilson loops is the fermionic $latitude$, which can be found from \eqref{DGRTc} by taking the curve to be a latitude on the $S^2$. Since a latitude is conformal to any other circle, the real feature of this loop are the scalar and fermionic couplings (with $q$ as in \eqref{superconn2})
\begin{equation}
\label{eq:matrixfermionic}
{M}_{I}^{J}=\begin{pmatrix}
 - q  & e^{-i \tau } 
   \sqrt{1-q^2} & 0 & 0 \\
e^{i \tau }  \sqrt{1-q^2}
   & q  & 0 & 0 \\
 0 & 0 & 1 & 0 \\
 0 & 0 & 0 & 1 \\
\end{pmatrix},\qquad
\eta_I^\alpha = \frac{1}{\sqrt{2}}\begin{pmatrix}\sqrt{1+q}\\ -\sqrt{1-q} e^{i\tau}\\0\\0 \end{pmatrix}_{I}     (1, -i e^{-i \tau})^\alpha,
\end{equation}
with $\tau\in [0,2\pi)$.

At the classical level, the fermionic latitude Wilson loop is cohomologically equivalent \cite{Bianchi:2018bke} to a linear combination of $bosonic$ latitudes\footnote{Loops in the fundamental representation for $U(N_1|N_2)$, $U(N_1)$ and $U(N_2$), respectively, are considered here.}
\be
\label{7:cohomoequiv}
 W^\text{fer}(q)= e^{-{i\pi q /2}} W_{\text{bos}}^{(1)}(q)-e^{{i\pi q /2}} W_{\text{bos}}^{(2)}(q)+Q(q)V,
 \ee
where $W_{\text{bos}}^{(1,2)}(q)$ are $bosonic$ latitude Wilson loops with scalar coupling governed by a matrix ${M}_{I}^{J}$ that coincides with (\ref{eq:matrixfermionic}) changing the last diagonal entry from $1$ to $-1$. In the above formula  $Q(q)$ is a linear combination of supercharges preserved by both bosonic and fermionic latitudes, while $V$ is a functional of the scalar fields and of the superconnection. Fermionic latitudes preserve 1/6 of the original supersymmetries and it is always possible to find two supercharges that do not depend on the parameters of the loops. The bosonic latitudes are instead 1/12 BPS and do not admit common preserved supercharges.
\subsection{Future directions}
A natural question that remains to be answered is of course if $any$ DGRT-like Wilson loop presented here is cohomologically equivalent to a bosonic Wilson loop, generalizing the bosonic circle constructed in \eqref{2:bosonic}. It would be also important to clarify the origin of the background term in the superconnection \eqref{superconn2}, a fact that seems a generic feature of fermionic loops, as already pointed out in Chapters~\ref{chapter:1} and \ref{chapter:2}.
\subsection{Acknowledgments}
It is a pleasure to thank L. Bianchi, M. Bianchi, V. Cardinali, N. Drukker, V. Forini, G. Martelloni, A. Mauri, S. Penati, M. Preti, D. Seminara, D. Trancanelli, E. Vescovi and I. Yaakov for collaborations, discussions and insights on many aspects of the subject presented. The work was supported by 
Istituto Nazionale di Fisica Nucleare (INFN) through the Gauge
and String Theory (GAST) research project.

\ifaffiliation
\subsection{Full affiliation}
Department of Mathematics, Physics and Computer Sciences, University of Parma and INFN Gruppo Collegato di Parma, 
Parco Area delle Scienze 7/A (Campus), 43124 Parma, Italy
\fi

\authors{Domenico Seminara\ifaffiliation, Universit\`a di Firenze and INFN Firenze\fi}
\section{The matrix model proposal for the latitude}
\label{chapter:8}
Among the family of Wilson operators on $S^2$ introduced in Chapter~\ref{chapter:7}, the latitudes 
are the simplest deformations of the circle. One may therefore hope that they are captured by a matrix model, as it occurs in $D=4$.
The fermionic latitude \eqref{eq:matrixfermionic} is 1/6 BPS and preserves $\osp(2|2)$, while 
the bosonic one (see Chapter~\ref{chapter:7}) is 1/12 BPS and only preserves $\mathfrak{u}(1|1)$. 
These two operators are cohomologically equivalent \eqref{7:cohomoequiv} as in the case of the two circle 
operators presented in Chapter~\ref{chapter:2}. Below we discuss a proposal for a matrix model evaluating both these latitudes.

\subsection{An educated guess for the matrix model}

The pattern of preserved supercharges and the analogies with the circle suggest that the expectation value of the bosonic latitude can also be computed in closed form using localization techniques.  
However the localization procedure has to yield a matrix model, which is a {\it significant} deformation of the one obtained in Chapter~\ref{chapter:3} and solved in Chapter~\ref{chapter:4}. In fact, an explicit three-loop computation of this observable in the fundamental representation (with $q$ defined in (\ref{superconn2}))
at framing 
$f$ in ABJM ($N_1=N_2=N$) gives
 \cite{Bianchi:2018bke}
\begin{equation}
\begin{aligned}
\label{chap8:boslat3loop}
\langle W^\text{bos}(q) \rangle_f &= 1+\frac{i \pi f N}{k}+\frac{\pi ^2}{6 k^2} \left[N^2 \left(3 (q ^2-f^2)+2\right)+1\right] 
\\&\quad{}
-\frac{i \pi ^3 N}{6 k^3} \left[N^2 \left(f^3+f \left(1-3 q ^2\right)+\left(q ^2-1\right) q \right)-4 f-q \left(q ^2-1\right)\right]+{\cal O}\left(k^{-4}\right).
\end{aligned}
\end{equation}
Analyzing this expression, we immediately realize that the $q$-dependence cannot be reabsorbed by a simple redefinition of the coupling constant at variance with the four-dimensional case. Therefore we expect that the deformation might affect both the measure of the matrix model and the observable that  we average to evaluate the Wilson loop. However, when we replace the observable with the identity the remaining matrix integral over the deformed measure must still give the partition function of ABJ(M) on $S^3$, namely the dependence of the partition function on $q$ must become trivial. We can perform an educated guess on the structure of this modification if we recall that the
original matrix model can be also viewed as a sort of supermatrix version of the partition function of Chern-Simons with gauge group $U(N_1+N_2)$ on $S^3/\mathbb{Z}_2$, where we have selected the vacuum that breaks the symmetry to $U(N_1)\times U(N_2)$ (see Chapter~\ref{chapter:4}).

A simple deformation enjoying this property is obtained by replacing $S^3$ with the squashed sphere $S^3_{\sqrt{ q}}$. Because of the topological nature of Chern-Simons, the partition function is unaffected by the squashing, up to framing anomalies. For ABJM this anomaly cancels, and we have the same partition function, while for ABJ they differ by a phase which is a polynomial in $(N_1-N_2)$ and cancels when we compute the average of the Wilson loop. 
 The modification amounts to replacing the original gauge contribution in the measure \eqref{eq:3:vector_multiplet_determinant} with
$
\prod_{i<j}^{N_i}4\sinh^2\frac{(\mu_{i}-\mu _{j})}{2}\mapsto
\prod_{i<j}^{N_i}4\sinh \frac{\sqrt{q}(\mu_{i}-\mu_{j})}{2}\sinh\frac{(\mu_{i}-\mu _{j})}{2\sqrt{q}} 
$
and similarly for the hypermultiplets. Namely we propose the following deformation \cite{Bianchi:2018bke} of the circle matrix model \eqref{eq:3:ABJ(M)_matrix_model} 
\begin{equation}
 \label{chap8:partitionf}
\begin{aligned}
 Z =\frac{1}{N_1!N_2!} &\int 
 \prod_{i=1}^{N_1}\frac{d\mu _{i}}{2\pi} \ e^{\frac{i k}{4\pi}\mu_{i}^{2}}
 \prod_{j=1}^{N_2}\frac{d\nu_{j}}{2\pi} \ e^{-\frac{i k}{4\pi}\nu_{j}^{2}}
\prod_{i=1}^{N_1}\prod_{j=1}^{N_2}\left(4\cosh\frac{\sqrt{q}(\mu _{i}-\nu_{j})}{2} 
\cosh\frac{(\mu _{i}-\nu_{j})}{2\sqrt{q}}\right)^{-1}
\\&\ {}\times
\prod_{i<j}^{N_1}4\sinh \frac{\sqrt{q}(\mu_{i}-\mu _{j})}{2}\sinh\frac{ (\mu_{i}-\mu _{j})}{2\sqrt{q}} 
\prod_{i<j}^{N_2}4\sinh\frac{\sqrt{q}(\nu_{i}-\nu_{j})}{2}  \sinh\frac{(\nu_{i}-\nu_{j})}{2\sqrt{q}} .
\end{aligned}
\end{equation}
This is the simplest non-trivial deformation of \eqref{eq:3:ABJ(M)_matrix_model} that lands back on the usual expression at $q=1$, and whose measure is symmetric under $q\leftrightarrow 1/q$. This symmetry is instrumental in recovering the correct conjectured 1/6 BPS $\theta$-Bremsstrahlung introduced in \cite{Correa:2014aga} (see Chapters~\ref{chapter:10} and~\ref{chapter:11}).
The expectation value of the bosonic Wilson latitude, for instance with a connection in the first gauge group, corresponds to the insertion in the matrix model of the quantity
$
\sum_{i=1}^{N_1} e^{\sqrt{q}\, \mu_{i}}.
$

\subsection{Comparison with the perturbative results}

We can check the proposal \eqref{chap8:partitionf} up to three loops against the perturbative result \eqref{chap8:boslat3loop} and similar results for ABJ presented in \cite{Bianchi:2018bke}. We find
perfect agreement if we assume that our field theory perturbative computation is performed at framing $q$. That the agreement manifests for this specific value of the framing is highly suggestive, since this is the precise value at which the conjectured cohomological equivalence with the fermionic Wilson loop is supposed to hold \eqref{7:cohomoequiv} and thus it allows us to use the matrix model results to reconstruct the expectation value of the fermionic latitude as well.

\subsection{Comparison with the string results}
The matrix model \eqref{chap8:partitionf} can be reformulated in terms of a Fermi gas. This representation provides a powerful tool for systematically expanding the partition function and Wilson loop observables in powers of $1/N$ at strong coupling. For simplicity we restrict the analysis to the ABJM slice, $N_1=N_2=N$. The final result can be expressed in terms of Airy functions and for the fermionic latitude takes a particular simple and elegant form
\begin{equation} \label{8:fermionicWL}
\langle W^\text{fer}(q) \rangle_q = -\frac{q \Gamma \left(-\frac{q}{2}\right) \text{Ai}\left(C^{-1/3} \left(N-B-2q/k\right)\right)}{2^{q+2} \sqrt{\pi }\, \Gamma \left(\frac{3-q}{2}\right)  \sin \left(2 \pi q /k\right)\text{Ai}\left(C^{-1/3} \left(N-B\right)\right)},
\end{equation}
where $C$ and $B$ have been defined right after \eqref{4:partitionfunction}. From this we can extract the leading contribution at large $N$
\begin{equation}\label{8:fermionicgenus0}
\langle W^\text{fer}(q) \rangle_q\big|_{g=0} = -i  \frac{ \Gamma \left(-\frac{q}{2}\right)}{2^{q +2} \sqrt{\pi} \, \Gamma \left(\frac{3-q}{2}\right)} e^{\pi q \sqrt{2 \lambda -{1}/{12}}}.
\end{equation}
 Classical string configurations that are dual to the fermionic latitude operators have been discussed in \cite{Correa:2014aga} and their leading exponential behavior scales according to $ \exp\big(\pi q\sqrt{2\lambda}\big)$ (see (\ref{WferGuille}) below),  which remarkably agrees with the expansion of the matrix model at strong coupling. Recently the one-loop string correction to the classical configuration were also computed in \cite{Medina-Rincon:2019bcc,David:2019lhr} and again perfect agreement with the matrix model was found (see Chapters~\ref{chapter:15} and \ref{chapter:12}).
\subsection{Future directions}
Despite the strong tests passed by the matrix model \eqref{chap8:partitionf}, it would be nice and instructive to have a complete derivation of it using localization. This might also help to understand if the other DGRT-like Wilson loops, which anyway share two supersymmetries with the fermionic latitude, can be evaluated in terms of a similar matrix model. In case of a positive answer, one might wonder whether an effective lower-dimensional theory describes this family of loops as it occurs in four dimensions.

\subsection{Acknowledgments}

The work presented here was carried out in collaboration with 
M. Bianchi, L. Griguolo, A. Mauri and S. Penati and was supported by 
Istituto Nazionale di Fisica Nucleare (INFN) through the Gauge
and String Theory (GAST) research project.

\ifaffiliation
\subsection{Full affiliation}
Dipartimento di Fisica e Astronomia, Universit\`a di Firenze and INFN, Sezione 
di Firenze, via G. Sansone 1, 50019 Sesto Fiorentino, Italy.
\fi

\authors{Silvia Penati\ifaffiliation, Universit\`a di Milano-Bicocca\fi~and Jiaju Zhang\ifaffiliation, SISSA\fi}

\section{Wilson loops in ${\cal N}=4$ supersymmetric Chern-Simons-matter theories}\label{chapter:9}

\subsection{General classification}
BPS Wilson loops in ${\mathcal N}=4$ supersymmetric Chern-Simons-matter (SCSM) theories were first studied in \cite{Ouyang:2015qma, Cooke:2015ila}. More general operators and an exhaustive classification for ${\mathcal N} \geq 2$ SCSM quiver theories were successively given in \cite{Ouyang:2015iza, Ouyang:2015bmy, Mauri:2017whf, Mauri:2018fsf}. Generalizing the operator of ABJM theory \eqref{2:bosonic}, a bosonic BPS operator which includes bi-linear couplings to scalars can be constructed for all theories with $2 \leq {\cal N} \leq 6$ supersymmetry. This is unique, up to R-symmetry rotations, and always preserves four real supercharges. More general classes of BPS operators with different amount of supersymmetry are obtained by introducing also couplings to fermions, in the spirit of \cite{Drukker:2009hy}.

In this chapter we review the general classification of BPS Wilson loops in ${\mathcal N}=4$ necklace quiver SCSM theories with gauge group and levels $\prod_{l=0}^{r-1}[U(N_{2l})_{-k}\times U(N_{2l+1})_{k}]$ \cite{Gaiotto:2008sd, Hosomichi:2008jd}. These theories are all conformal and can be obtained by a quotient of the $U(N)_k \times U(M)_{-k}$ ABJ theory where we decompose $N = N_1 + N_3 + ... + N_{2r-1}$ and $M = N_{0} + N_{2} +  ... + N_{2r-2}$. They have a string dual description in terms of M-theory on the orbifold background $AdS_4 \times S^7/(\mathbb{Z}_r \oplus \mathbb{Z}_r)/\mathbb{Z}_k$. When $N_{0} = ... = N_{2r-1}$ they reduce to the $\mathbb{Z}_r$-orbifold of ABJM \cite{Benna:2008zy} which is dual to M-theory on $AdS_4 \times S^7/(\mathbb{Z}_r \oplus \mathbb{Z}_{rk})$.

A large class of fermionic BPS Wilson loops in ${\mathcal N}=4$ SCSM theories can be obtained by the orbifold decomposition of fermionic 1/6 BPS and 1/2 BPS operators of the ABJM or ABJ theories \cite{Mauri:2017whf}. For circular contours this leads in general to 1/4 BPS fermionic operators
corresponding to superconnections of the form  (for $r \geq 3$)
\begin{equation} \label{9:sc1}
{\mathcal L} = \begin{pmatrix}
{\mathcal A}^{(1)}    & f_1^{(1)}  & h_1^{(1)} & 0         & \cdots       \\
f_2^{(1)}    & {\mathcal A}^{(2)}  & f_1^{(2)} & h_1^{(2)} & \ddots       \\
h_2^{(1)}    & f_2^{(2)}  & {\mathcal A}^{(3)} & f_1^{(3)} & \ddots       \\
0            & h_2^{(2)}  & f_2^{(3)} & {\mathcal A}^{(4)} & \ddots       \\
\vdots       & \ddots     & \ddots    & \ddots             & \ddots
\end{pmatrix}.
\end{equation}
Here the diagonal blocks ${\mathcal A}^{(l)} $ contain the usual gauge and scalar field couplings, the next-to-diagonal blocks $f_{1,2}^{(l)}$ are linear in fermions, mimicking the entries of \eqref{2:superconnection} for ABJM operators, whereas the next-to-next-to-diagonal blocks $h_{1,2}^{(l)}$ are quadratic expressions in the scalars. For the case of orbifold ABJM theory their explicit expressions can be found in \cite{Mauri:2017whf}, as functions of several complex parameters.
Up to R-symmetry rotations, the corresponding Wilson loops have been classified into two independent classes, where each class is parametrized by three complex moduli \cite{Mauri:2017whf}.\footnote{{The other two classes found in \cite{Ouyang:2015iza,Ouyang:2015bmy,Mauri:2017whf} are equivalent to the bosonic Wilson loop. We thank N.~Drukker for pointing it out.}}
Wilson loops belonging to different classes differ primarily by the chirality of the fermionic couplings. The classification of the corresponding preserved supercharges reveals a high degree of degeneracy: The operators of the {two} classes share the same set of supersymmetries.
More general 1/4 BPS Wilson loops can be constructed, which are not obtained by orbifold decompositions of ABJM operators.
These operators fall outside the aforementioned classes, but are block-diagonalizable to them.

For special choices of the parameters the superconnection in \eqref{9:sc1} becomes block-diagonal and the corresponding Wilson loop reduces to a sum $W = \sum _{l=1}^r W^{(l)}$, where the ``double-node'' $W^{(l)}$ is the holonomy of a $U(N_{l-1}|N_{l})$ superconnection of the type \eqref{2:superconnection}, which includes $(N_{l-1}, N_{l})$ gauge fields and matter coupled to them.
{In each of the two classes, for a particular set of parameters the preserved supersymmetry gets enhanced and we obtain fermionic 1/2 BPS Wilson loops \cite{Ouyang:2015qma,Cooke:2015ila,Mauri:2017whf}. In the notation of \cite{Cooke:2015ila} we refer to the corresponding double-node operators as $\psi_1$-loop and $\psi_2$-loop, respectively.}

All fermionic 1/4 and 1/2 BPS operators are in the same {$Q$}-cohomological class of the bosonic 1/4 BPS Wilson loop, where {$Q$} is a conserved supercharge shared by all the operators. Therefore, they are in principle amenable of exact evaluation via the matrix model that computes the bosonic operator \cite{Kapustin:2009kz, Marino:2012az}.

\subsection{Degenerate Wilson loops}

As already mentioned, fermionic 1/4 BPS Wilson loops belonging to different classes preserve the same set of supercharges.
{In particular, this occurs for the two kinds of 1/2 BPS Wilson loops, $\psi_1$-loop and $\psi_2$-loop.}
In the orbifold ABJM case these two operators come from quotienting two 1/2 BPS Wilson loops of the ABJM theory that share eight real supercharges. Therefore, the total degeneracy appearing in the ${\cal N}=4$ theory can be understood as the legacy of the partial overlapping of conserved supercharges already present in the parent ABJM theory. In a Higgsing construction, $\psi_1$- and $\psi_2$-loops correspond to exciting non-relativistic infinitively massive particles or antiparticles, respectively \cite{Lietti:2017gtc}.

The degeneracy of 1/2 BPS Wilson loops opens important questions. Which are their gravity duals? Do we expect degeneracy also in the corresponding M2-brane solutions or is the actual BPS Wilson loop that survives at strong coupling a linear combination of operators, as first suggested in
\cite{Cooke:2015ila}? Second, which is the fate of this degeneracy at quantum level? Do degenerate operators share the same expectation value and how does this expectation value match the matrix model prediction?

For the orbifold ABJM theory, the first question was answered in \cite{Lietti:2017gtc} by identifying {the degeneracy of} $\psi_1$- and $\psi_2$-loops {with the degeneracy of} a pair of M2 and anti-M2-branes localised at different positions in the compact space, and preserving the same set of supercharges.
The second question has been addressed in \cite{Bianchi:2016vvm, Griguolo:2015swa} for theories with groups of unequal  ranks. Assuming that the classical cohomological equivalence is compatible with the localization procedure, the matrix model predicts $\langle W_{\psi_1} \rangle = \langle W_{ \psi_2} \rangle$ and its exact expression expanded at weak coupling and at framing zero exhibits vanishing contributions at odd orders. However, in \cite{Griguolo:2015swa} it was shown that at three loops, at least in the three-node color sector, $\langle W_{\psi_1} \rangle|_{\lambda^3}  = -\langle W_{\psi_2} \rangle|_{\lambda^3} \neq 0$. This implies that in theories with unequal group ranks only the linear combination $\langle W_{\psi_1} +W_{\psi_2} \rangle$ can match the matrix model prediction, pointing towards a non-trivial uplifting of the classical degeneracy.

\subsection{Future directions}

At the moment there are no exhaustive answers to the previous questions. In the orbifold ABJM theory the degeneracy is not broken at strong coupling, in line with the matrix model prediction, but a confirmation from a genuine perturbative calculation is still lacking.
In the more general case of theories with unequal ranks, the matching with the matrix model implies an uplifting of degeneracy, but it would be important to find the dual M2-brane configurations to have confirmation at strong coupling.
Moreover, a similar analysis should be extended to degenerate fermionic 1/4 BPS Wilson loops for which the dual configurations are not known. {In particular, for generic parametric dependent operators in the two classes, a perturbative calculation would provide parametric dependent expectation values \cite{Mauri:2018fsf}, but there is no correspondingly free parameter in the matrix model prediction. Moreover, if the gravity duals of all of these degenerate operators exist as different brane configurations}, it would be interesting to understand how to flow in the moduli space from one brane configuration to another. This is another problem that deserves further investigation.

Finally, similar configurations of degenerate Wilson loops occur also in $\mathcal N<4$ SCSM theories, where the problem of identifying the corresponding gravity duals and matching the matrix model predictions with their expectation values \cite{Mauri:2018fsf} is still to be fully addressed.

\subsection{Acknowledgements}
We thank M.~S.~Bianchi, L.~Griguolo, M.~Leoni, M.~Lietti, A.~Mauri, H.~Ouyang, D.~Seminara and J.-B.~Wu for their collaboration on the subject of this session. This work has been supported in part by the Ministero dell'Istruzione, Universit\`a e Ricerca (MIUR) of Italy and by the Istituto Nazionale di Fisica Nucleare (INFN) through the ``Gauge Theories, Strings, Supergravity'' (GSS) research projects.

\ifaffiliation
\subsection{Full affiliation}
Silvia Penati, Dipartimento di Fisica, Universit\`a degli Studi di Milano-Bicocca, \& INFN, Sezione di Milano-Bicocca, Piazza della Scienza 3, I-20126 Milano, Italy.
\\Jiaju Zhang, SISSA \& INFN Sezione di Trieste, Via Bonomea 265, 34136 Trieste, Italy
\fi

\authors{Michelangelo Preti\ifaffiliation, Nordita\fi}
\section{Bremsstrahlung functions I: Definition and perturbative results}\label{chapter:10}

\subsection{The generalized cusp and the Bremsstrahlung functions in ABJ(M)}

\begin{figure}[!h]
  \begin{center}
  \includegraphics[width=12cm]{\folder/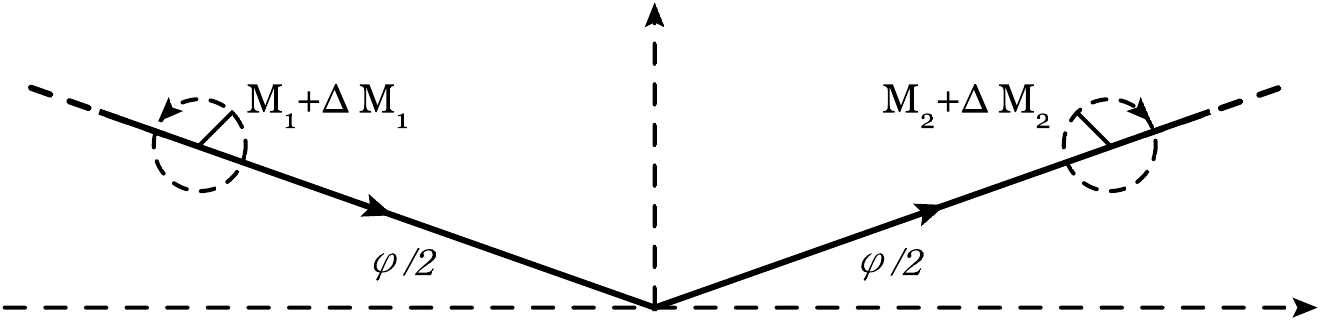}
  \caption{The planar Euclidean cusp with angular opening $\pi-\varphi$ between the Wilson lines parametrized by $x^\mu=\{\tau \cos \varphi/2,|\tau| \sin \varphi/2,0\}$ with $-\infty\leq\tau\leq\infty$. The operators lying on it possess also a discontinuity in the R-symmetry space represented by the different orientations of the matter couplings $M_a+\Delta M_a$, with $a=1,2$.}
  \label{10:cusp}
  \end{center}
\end{figure}

The bosonic and fermionic Wilson operators \eqref{2:bosonic} and \eqref{2:loop} can be also supported along infinite lines. 
In this case, the constant piece $|\dot{x}|/4|x|$ introduced in the fermionic superconnection \eqref{2:loop} disappears, while the couplings to the matter become constant. 

When a cusp with angle $\varphi$ is introduced into the 1/2 and 1/6 BPS lines, as in Figure~\ref{10:cusp}, supersymmetry is completely broken and the expectation value of the Wilson operator develops a divergence. 
The coefficient of the divergence can be analysed in very general terms \cite{Korchemsky:1987wg} and is called {\it cusp anomalous dimension}. Moreover, one can introduce a second deformation by an internal angle $\theta$ affecting the scalar and fermionic couplings (the latter are fully fixed by the scalar ones as in \eqref{2:eta}) such that
\begin{equation}
\Tr [(M_a+\Delta M_a) (M_b+\Delta M_b)] =
     \begin{cases}
       4 \cos^2\frac{\theta}{2} &\quad a\neq b\,,\\
       4                                   &\quad a=b\,,
     \end{cases},
     \end{equation}
where the indices $a,b=1,2$ represent the two sides of the cusp contour (see Figure~\ref{10:cusp}) and $\Delta M$ vanishes in the bosonic case.
Then the expectation value of the Wilson operators can be written as
\begin{equation}\begin{split}\label{10:Gammacusp}
 \log \langle W_{\text{cusp}}^{\text{bos}}\rangle\sim-\Gamma_{1/6}(k,N_1,N_2,\varphi,\theta) \log{L}/{\epsilon}
 \qquad
 \log \langle W_{\text{cusp}}^{1/2}\rangle\sim-\Gamma_{1/2}(k,N,\varphi,\theta) \log{L}/{\epsilon},
\end{split}\end{equation}
where $L$ and $\epsilon$ are the IR and UV regulators, respectively. Since the cusp anomalous dimension for the cusp with 1/2 BPS rays is only known for equal ranks of the gauge group, we set $N_1=N_2=N$ for this quantity.
The coefficients  $\Gamma_{1/6}$ and $\Gamma_{1/2}$ of the logarithm depend on both angles and are called {\it generalized cusp anomalous dimensions} \cite{Griguolo:2012iq,Correa:2014aga}. 
$W^{1/2}_\text{cusp}$ preserves two supercharges when $\varphi^2=\theta^2$, 
while $W^{1/6}_\text{cusp}$ is only BPS for $\varphi=\theta=0$. 
As a consequence, for small angles the cusp anomalous dimensions take the form
\begin{equation}\label{10:brem12}
 \Gamma_{1/6}\sim \theta^2B_{1/6}^\theta(k,N_1,N_2)-\varphi^2B_{1/6}^\varphi(k,N_1,N_2)\,,
 \qquad
  \Gamma_{1/2}\sim (\theta^2-\varphi^2) B_{1/2}(k,N)\,.
\end{equation}
The $B$'s are known as the {\it Bremsstrahlung functions}. In a conformal field theory, these functions also govern the energy radiated by an accelerating massive probe \cite{Correa:2012at}, hence the name.

\subsection{Renormalization and perturbation theory}

Following \cite{Korchemsky:1987wg}, the cusp anomalous dimension is extracted from 
$\langle W_{\text{cusp}}\rangle$. First one subtracts the IR gauge-dependent divergences by introducing a multiplicative renormalization $Z_\text{open}$, which is equivalent to the subtraction of the straight line. 
For $\langle W_{\text{cusp}}^{\text{bos}}\rangle$ this term vanishes, so the renormalized Wilson loops are
\begin{equation}
\langle W_{\text{cusp}}^{\text{bos}} \rangle_\text{ren}=Z^{-1}_{1/6}\langle W_{\text{cusp}}^{\text{bos}}  \rangle\quad\text{and}\quad
\langle W_{\text{cusp}}^{1/2} \rangle_\text{ren}=Z^{-1}_{1/2}Z^{-1}_\text{open}\langle W_{\text{cusp}}^{1/2}  \rangle.
\end{equation}
$\Gamma$ arises from the renormalization group equations for the anomalous dimensions of the non-local operators
\begin{equation}\label{10:gammadefZ}
\Gamma_{1/6}=\frac{d \log Z_{1/6}}{d\log \mu}\quad\text{and}\quad
\Gamma_{1/2}=\frac{d \log Z_{1/2}}{d\log \mu},
\end{equation}
where the derivative is taken with respect to the renormalization scale $\mu$. 

Given $\Gamma$, it is possible to compute the Bremsstrahlung functions using  \eqref{10:brem12}
\begin{equation}\label{10:bremder}
B_{1/2}=\frac12\frac{\partial^2\Gamma_{1/2}}{\partial \theta^2}\biggl|_{\varphi,\theta=0}=-\frac12\frac{\partial^2\Gamma_{1/2}}{\partial \varphi^2}\biggl|_{\varphi,\theta=0},\quad
B^\theta_{1/6}=\frac12\frac{\partial^2\Gamma_{1/6}}{\partial \theta^2}\biggl|_{\varphi,\theta=0}\;\text{and}\quad
B^\varphi_{1/6}=-\frac12\frac{\partial^2\Gamma_{1/6}}{\partial \varphi^2}\biggl|_{\varphi,\theta=0}.
\end{equation}

The $Z$'s can be evaluated in perturbation theory in dimensional regularization $d=3-2\epsilon$ as in Chapter~\ref{chapter:5}. 
The Feynman diagrams providing an expansion in the coupling $1/k$ and 
the relevant integrals can be performed directly in $x$-space, solving first the internal integrations and then the ones on the Wilson line contour. A more efficient strategy at higher loops is to Fourier transform the integrals to momentum space and perform the contour integration first. Using this procedure, the integrals resemble those of non-relativistic Feynman integrals arising in the heavy quarks effective theory (HQET) \cite{Grozin:2014hna,Grozin:2015kna}. 
Finally, using \eqref{10:gammadefZ}, it is possible to extract the anomalous dimensions from the residues of the simple poles in $\epsilon$ of $Z_{1/6}$ and $Z_{1/2}$. 

In the following we summarize the main perturbative results at weak-coupling for these functions. The strong coupling expansions are presented in Chapter~\ref{chapter:12}.

\subsection{Weak coupling expansion of the Bremsstrahlung function for the  fermionic cusp}

The cusp anomaly $\Gamma_{1/2}(k,N,\varphi,\theta)$ was computed at two-loops via perturbation theory in \cite{Griguolo:2012iq}.
In the limit in which only ladder diagrams contribute, it is known exactly by resumming the perturbative series with the Bethe-Salpeter method  \cite{Bonini:2016fnc}. The case $\varphi=0$ was explored at three-loops using the HQET formalism in \cite{Bianchi:2017svd, Preti:2017fjb}. Using \eqref{10:bremder} one obtains
\begin{equation}
\label{10:B12perturbative}
B_{1/2}(k,N)=\frac{N}{8k}-\frac{\pi^2N(N^2-3)}{48k^3}+\mathcal{O}\!\left(k^{-5}\right).
\end{equation}
The computation suggests that $B_{1/2}$ has an expansion in odd powers of the coupling. This fact is confirmed by the exact computation in terms of  multiple-wound Wilson circles (see Chapter~\ref{chapter:11}).

\subsection{Weak coupling expansion of the Bremsstrahlung functions for the bosonic cusp}

The function $B_{1/6}^{\varphi}$ associated to the small angle limit of the geometric bosonic cusp anomaly was computed using \eqref{10:bremder} in \cite{Griguolo:2012iq,Bianchi:2014laa} and it is given by 
\begin{equation}\label{10:B1}
B_{1/6}^\varphi(k,N_1,N_2)=\frac{N_1N_2}{2k^2}+\mathcal{O}\!\left(k^{-4}\right).
\end{equation}
This result at equal gauge group ranks coincides with the proposed exact formula in \cite{Lewkowycz:2013laa}.

The function $B^{\theta}_{1/6}$ corresponding to a cusp in R-symmetry space along a 1/6 BPS straight line ($\varphi=0$) is computed at two-loop in \cite{Bianchi:2014laa} and at higher order in \cite{Bianchi:2017afp,Bianchi:2017ujp}, leading to
\begin{equation}\label{10:B2}
B_{1/6}^\theta(k,N_1,N_2)=\frac{N_1N_2}{4k^2}-\frac{\pi^2N_2(5{N_1}^2N_2+N_1{N_2}^2-3N_1-5N_2)}{24k^4}+\mathcal{O}\!\left(k^{-6}\right),
\end{equation}
for generic ranks of the gauge groups. This result is compatible with the exact computation via defect theory (see Chapter~\ref{chapter:11}) and the bosonic latitude matrix model proposal (see Chapter~\ref{chapter:8}).

As expected, both \eqref{10:B1} and \eqref{10:B2} have an expansion in even power of the coupling. Indeed, ABJ(M) Wilson loops with planar contours computed at framing zero (see Chapter~\ref{chapter:6}) automatically have vanishing expectation values at odd loops \cite{Rey:2008bh}.

\subsection{Future directions}

In analogy with the four-dimensional case \cite{Gromov:2012eu,Bonini:2015fng}, it could be interesting to extend the definitions of the Bremsstrahlung functions by adding $L$ units of R-charge. This could make them accessible from both integrability and localization techniques.
Another possible future direction is the evaluation of the cusp anomalous dimension and its small angle limit for the cusp with 1/2 BPS rays in the case of generic ranks. This study could shed some light on the exponentiation property of the fermionic Wilson loops in ABJ.
It would be interesting to extend the analyses of Bremsstrahlung functions to the operator with fermionic 1/6 BPS rays.

\subsection{Acknowledgements}
We thank L. Bianchi, M. S. Bianchi, M. Bonini, L. Griguolo, A. Mauri, S. Penati and D. Seminara for their collaboration on \cite{Bonini:2016fnc,Bianchi:2017svd}, on which this chapter is based.

\ifaffiliation
\subsection{Full affiliation}
Nordita, KTH Royal Institute of Technology and Stockholm University\\
Roslagstullsbacken 23, SE-10691 Stockholm, Sweden
\fi
\authors{Lorenzo Bianchi\ifaffiliation, Queen Mary University of London\fi}
\section{Bremsstrahlung functions II: Nonperturbative methods}\label{chapter:11}
\subsection{Wilson lines as superconformal defects}
Since the cusped Wilson line does not preserve any supersymmetry, one may expect this would kill any hope of using supersymmetric localization. Nevertheless, for small angles one can relate the cusp anomalous dimension to conformal defect correlation functions \cite{Correa:2012at}, leading to exact results. A supersymmetric Wilson line breaks translational as well as R-symmetry leading to the associated Ward identities
\begin{equation}\label{11:dertmn}
\partial_{\mu}T^{\mu \bar{\mu}}=\delta^2(x_{\perp}) \mathbb{D}^{\bar{\mu}}(\tau)\,,
\qquad
\partial_{\mu}j^{\mu\mathcal{I}}=\delta^2(x_{\perp}) \mathbb{O}^\mathcal{I}(\tau)\,,
\end{equation}
where the delta function localizes the r.h.s. on the defect profile (a straight line along the direction 1 in this case), $\bar{\mu}=2,3$ label orthogonal directions and the index $\mathcal{I}$ spans the subset of R-symmetry generators that are broken by the defect. The defect excitation $\mathbb{D}^{\bar{\mu}}$ is usually called \emph{displacement operator}. The bosonic Wilson line \eqref{2:bosonic} preserves $\su(1,1|1)\oplus \su(2) \oplus \su(2)\subset \osp(6|4)$, thus breaking 8 of the 15 $SU(4)_R$ generators. We label the associated defect operators as $\mathbb{O}^{a\dot a}$ and $\bar{\mathbb{O}}^{\dot a a}$, with $a$ and $\dot a$ fundamental indices for the preserved R-symmetry $SU(2)\times SU(2)\times U(1)$\cite{Bianchi:2018scb}. The fermionic Wilson line \eqref{2:loop}, instead, preserves $\su(1,1|3)\oplus \mathfrak{u}(1) \subset \osp(6|4)$, breaking only 6 generators. The associated defect operators are organised in fundamental $\mathbb{O}^A$, and antifundamental $\bar{\mathbb{O}}_A$ representations of the preserved $SU(3)$. It is worth stressing that equations \eqref{11:dertmn} are written in a loose notation and must be interpreted as a Ward identity when both sides are inserted inside a correlation function with other operators. In particular, for the fermionic case, the natural object to be inserted on the Wilson line is a $U(N|N)$ supermatrix and both the displacement and the R-symmetry operators admit an explicit realization in terms of supermatrices \cite{Bianchi:2017ozk}.

Considering a generalized cusp deformation one finds \cite{Correa:2014aga,Bianchi:2017ozk}
\begin{align}
 &\text{Bosonic} & \langle\!\langle\mathbb{D}^{\bar\mu}(\tau)\mathbb{D}^{\bar\nu}(0)\rangle\!\rangle_{\text{bos}}&=12 B^{\varphi}_{1/6} \frac{\delta^{\bar\mu\bar\nu}}{|\tau|^4}, & \langle\!\langle{\mathbb{O}}^{a\dot a}(\tau) \bar{\mathbb{O}}^{\dot b b}(0)\rangle\!\rangle_{\text{bos}}&=-4 B^{\theta}_{1/6} \frac{\epsilon^{ab} \epsilon^{\dot a\dot b}}{|\tau|^2}, \\
 &\text{Fermionic} &\langle\!\langle\mathbb{D}^{\bar\mu}(\tau)\mathbb{D}^{\bar\nu}(0)\rangle\!\rangle_{\text{ferm}}&= 12 B_{1/2} \frac{\delta^{\bar\mu\bar\nu}}{|\tau|^4}, & \langle\!\langle{\mathbb{O}}^{A}(\tau) \bar{\mathbb{O}}_{B}(0)\rangle\!\rangle_{\text{ferm}}&=-4 B_{1/2} \frac{\delta^A_B}{|\tau|^2} .
\end{align}
The smaller amount of preserved supersymmetry naively prevents one from relating $B^{\varphi}_{1/6}$ and $B^{\theta}_{1/6}$ using defect supersymmetric Ward identities (in this case $\mathbb{D}$ and $\mathbb{O}$ do not belong to the same supermultiplet as it happens for the 1/2 BPS case). Nevertheless, it was shown in \cite{Bianchi:2018scb} that a class of vanishing three-point functions allows to write Ward identities with broken supercharges giving a formal derivation of the relation
\begin{align}
 B^{\varphi}_{1/6}=2 B^{\theta}_{1/6}.
\end{align}

\subsection{Relation to circular Wilson loops}
Using the fact that defect two-point functions are the same for the straight line and the circular case, one can relate the Bremsstrahlung functions to specific deformations of the circular Wilson loop. Specifically, for the latitude bosonic and fermionic Wilson loops (see Chapter~\ref{chapter:7}) one can prove \cite{Correa:2014aga,Bianchi:2017ozk}
\begin{equation}\label{11:dernu}
B^{\theta}_{1/6}=\frac{1}{4\pi^2} \Big. \partial_q \log |\langle W^{\text{bos}} (q) \rangle | \Big|_{q=1} \,,
\qquad
B_{1/2}=\frac{1}{4\pi^2} \Big. \partial_q \log \langle W^{\text{fer}} (q) \rangle \Big|_{q=1}\,.
\end{equation}
For the fermionic case, this relation was previously conjectured in \cite{Bianchi:2014laa}. The problem with these relations for ABJM theory is that no first-principle localization calculation exists for the latitude Wilson loops (see, however, Chapter~\ref{chapter:8} of this review for a proposal in this direction \cite{Bianchi:2018bke}). Furthermore, to extend the localization results to the fermionic case one has to rely on the cohomological equivalence (see Chapter~\ref{chapter:3}), which involves subtleties associated with non-integer framing (see Chapter~\ref{chapter:6})\cite{Bianchi:2014laa,Bianchi:2016yzj}. Despite these difficulties one can still achieve exact expressions for the Bremsstrahlung functions as we discuss in the following.

\subsection{Exact Bremsstrahlung for the bosonic Wilson loop}
\begin{figure}[h]
\begin{center}
\includegraphics[width=.45\textwidth]{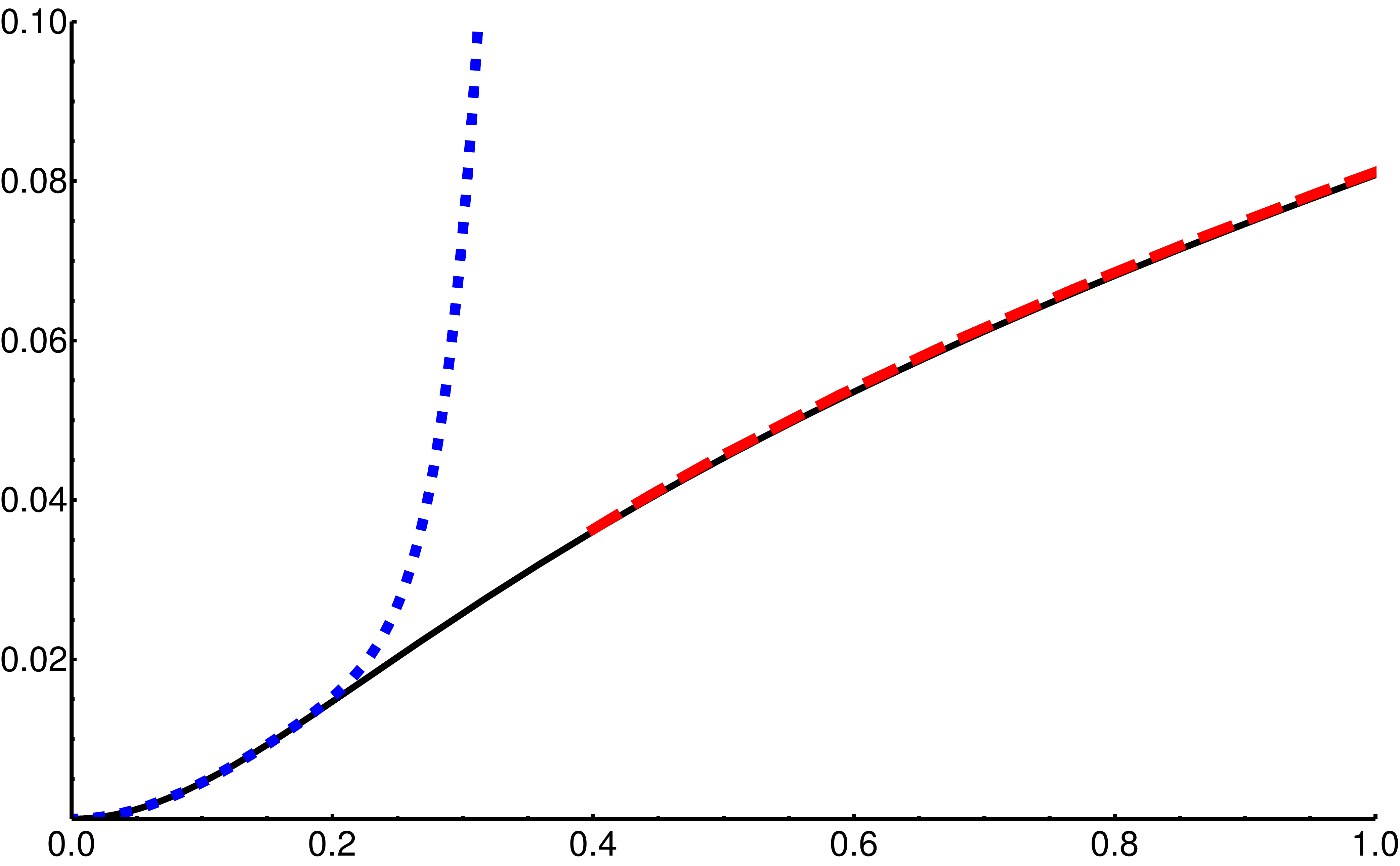}
\put (-78,50) {$B_{1/6}$}
\put (1,2) {$\lambda$}
\qquad
\includegraphics[width=.45\textwidth]{\folder/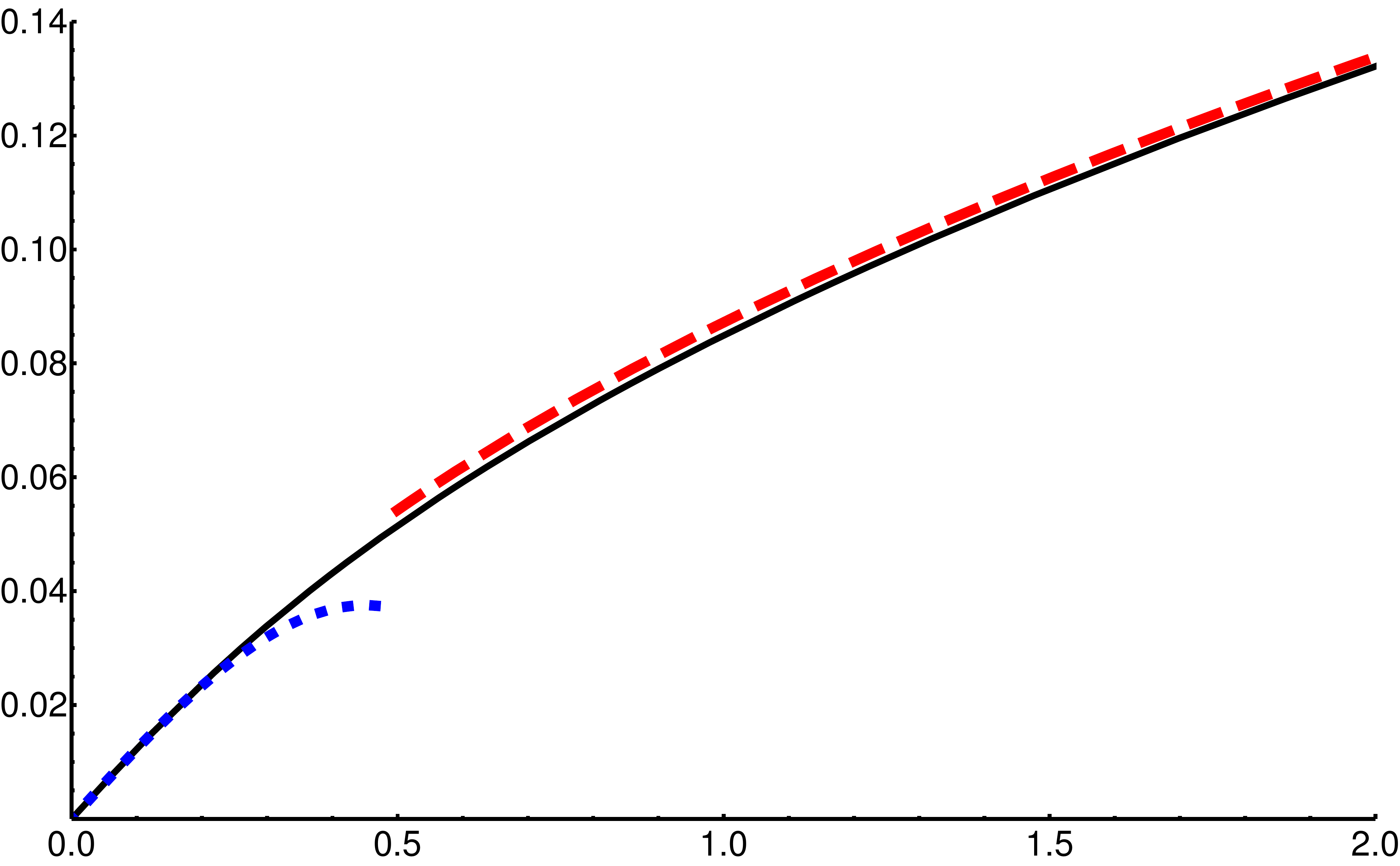}
\put (-78,50) {$B_{1/2}$}
\put (1,2) {$\lambda$}
\caption{Left: The Bremsstrahlung function for the bosonic Wilson loop at large $N$ computed using \eqref{11:derm}. It interpolates nicely between the weak coupling (blue, dotted) and strong coupling (red, dashed) expansions.\\
Right: The Bremsstrahlung function for the fermionic Wilson loop at large $N$, given by \eqref{11:B12exact}. It interpolates nicely between the weak coupling (blue, dotted) and strong coupling (red, dashed) expansions.}
\label{11:fig}
\end{center}
\end{figure}

An exact expression for $B^{\varphi}_{1/6}$ appeared already in \cite{Lewkowycz:2013laa}, based on the conjectured relation with the stress tensor one-point function
\begin{equation}\label{11:defh}
B^{\varphi}_{1/6}= 2a_{T}, 
\qquad
\langle T^{11}(x_{\perp})\rangle_{\text{bos}}=\frac{a_T}{|x_{\perp}|^3},
\end{equation}
which was later shown, in a slightly different context, to be a consequence of supersymmetric Ward identities \cite{Bianchi:2018zpb}. The convenient feature of equation \eqref{11:defh} is that the stress tensor one-point function can be computed by supersymmetric localization. In particular, exploiting the definition of the supersymmetric R\'enyi entropy of \cite{Nishioka:2013haa}, the authors of \cite{Lewkowycz:2013laa} showed that $a_T$ can be computed by
\begin{align}\label{11:derm}
 a_T=\frac{1}{8\pi^2} \partial_m \log \langle W_m \rangle \Big |_{m=1},
\end{align}
where $W_m$ is a circular Wilson loop winding $m$ times \cite{Marino:2009jd,Klemm:2012ii}. Despite no closed form expression is available for the r.h.s. of \eqref{11:derm}, it is not too hard to expand it at weak and strong coupling, or evaluate it numerically at finite coupling (see Figure~\ref{11:fig} (left) for a plot and \cite{Lewkowycz:2013laa} for further details).

\subsection{Exact Bremsstrahlung for the fermionic Wilson loop}
The first proposal for the fermionic Bremsstrahlung function appeared in \cite{Bianchi:2014laa}, based on several assumptions related to the cohomological equivalence and the dependence on the framing, which were then clarified in various papers \cite{Bianchi:2014laa,Bianchi:2016yzj,Bianchi:2017ozk,Bianchi:2018scb,Bianchi:2018bke}. The upshot is that, combining the relations \eqref{11:dernu}, \eqref{11:defh} and \eqref{11:derm} it is possible to establish a connection between the matrix model for the winding Wilson loop and the matrix model for the geometric deformation $\nu$. This leads to interesting relations between bosonic and fermionic Bremsstrahlung functions as well as, notably, to a closed form expression for $B_{1/2}$, which was derived in \cite{Bianchi:2018scb} and we present here in a new and simpler form for the large $N$ case
\begin{equation}\label{11:B12exact}
B_{1/2}=\frac{\kappa}{64 \pi }  \, _2F_1\left(\frac{1}{2},\frac{1}{2};2;-\frac{\kappa ^2}{16}\right),
\qquad
\lambda=\frac{\kappa}{8 \pi }  \, _3F_2\left(\frac{1}{2},\frac{1}{2},\frac{1}{2};1,\frac{3}{2};-\frac{\kappa ^2}{16}\right),
\end{equation}
where the effective coupling $\kappa$ was already defined in \eqref{4:kappa}. In Figure~\ref{11:fig} (right) we plot this function together with the weak and strong coupling expansions.

\subsection{Future directions}
A natural future direction would be to obtain an expression of the ABJM Bremsstrahlung function using integrability, along the lines of the $\mathcal{N}=4$ SYM result \cite{Drukker:2012de,Correa:2012hh}. This would lead to an honest derivation of the interpolating function $h(\lambda)$ (see Chapter~\ref{chapter:13}). Another interesting avenue to explore is the study of higher points correlation functions of defect operators. The defect theory provides a tractable example of 1d CFT with an interesting $AdS_2$ dual \cite{Giombi:2017cqn}.

\subsection{Acknowledgements}
It is a pleasure to thank L. Griguolo, D. Seminara, M. Preti and E. Vescovi for their collaboration on the works \cite{Bianchi:2017ozk,Bianchi:2018scb}. This work is supported by the European Union’s Horizon 2020
research and innovation programme under the Marie
Sk{\l}odowska-Curie grant agreement No 749909.

\ifaffiliation
\subsection{Full affiliation}
Center for Research in String Theory - School of Physics and Astronomy\\
Queen Mary University of London, Mile End Road, London E1 4NS, UK
\fi
\authors{Guillermo A. Silva\ifaffiliation, Universidad Nacional de La Plata\fi}
\section{Holography for ABJM Wilson loops I: Classical strings }
\label{chapter:15}

\subsection{Holographic duals of ABJM theory}

ABJM theory is dual to M-theory on $AdS_4\times S^7/\mathbb Z_k$. In the large $k$ limit, the theory reduces to type IIA string theory on $AdS_4 \times \mathbb{CP}^3$. Sticking to the string picture, the IIA background comprises 
\begin{equation}
ds^2=L^2(ds^2_{\sf AdS_4}+4ds^2_{\mathbb {CP}^3}),
\qquad
e^{2\phi}=4\frac{L^2 }{k^2 },
\qquad
F^{(4)}=\frac{3}{2}kL^2\,\text{vol}({AdS_4}),
\qquad
F^{(2)}=\frac{k}{4}dA,
\label{15:IIA}
\end{equation}
with $A=\cos{\alpha}\,d\chi + 2\cos^2{\frac{\alpha}{2}}\cos{\theta_1}d\varphi_1 + 2\sin^2{\frac{\alpha}{2}}\cos{\theta_2}d\varphi_2$ and the $\mathbb {CP}^3$ metric written as
\begin{equation}
\begin{aligned}
ds^{2}_{\mathbb{CP}^3} =
\frac{1}{4}\Bigl[&d\alpha^2+\cos^2{\frac{\alpha}{2}}
\left(d\theta^2_1+\sin^2{\theta_1} d\varphi^2_1\right)+\sin^2{\frac{\alpha}{2}}\left(d\theta^2_2+\sin^2{\theta_2}d\varphi^2_2\right)
\\&{}+\sin^2{\frac{\alpha}{2}}\cos^2{\frac{\alpha}{2}}\left(d\chi+\cos{\theta_1}d\varphi_1-\cos{\theta_2}d\varphi_2\right)^2\Bigr]\,.
\label{15:cp3}
\end{aligned}
\end{equation}
The relation between IIA string theory and ABJM parameters   in the 't~Hooft limit is: $ {L^2}/{\alpha'}=\pi\sqrt{2\big(\lambda- \frac1 {24}\big)}$ and $ g_s^2= \pi {(2\lambda)^{5/ 2}}/{N^2}$.

ABJM Wilson loops in the fundamental representation map to IIA open strings partition functions. The leading order contribution at strong coupling arises from minimal surfaces. We expect the specific boundary conditions for the string worldsheet to be dictated by the Wilson loop data ${\cal C}, M$ and $\eta$.

\subsection{The known}

\begin{figure}[h]
\centering
\includegraphics[width=\textwidth]{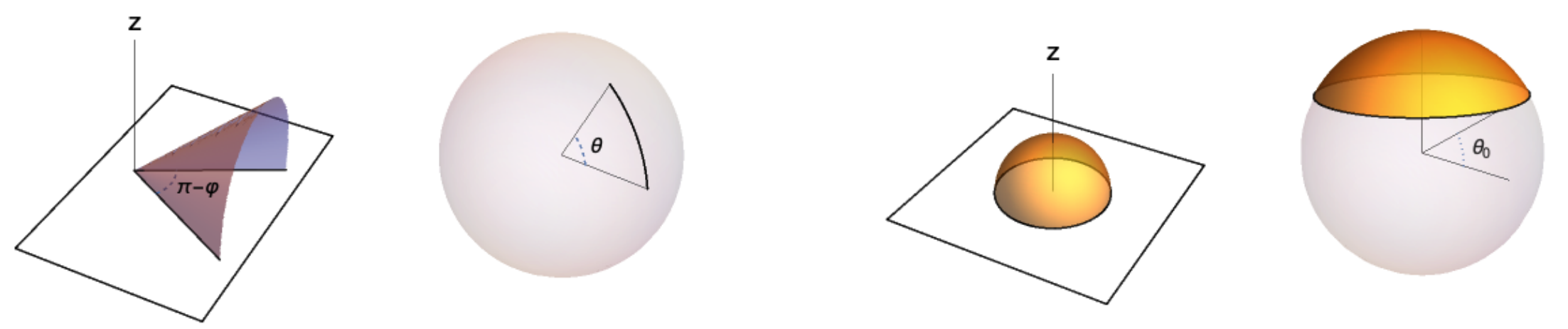}
\caption{
Left: The generalized cusp's dual worldsheet sits on the geometrical cusp $\varphi$ at $z=0$ and spans a great  arc of amplitude $\theta$ in $\mathbb{CP}^3$. \\
Right: String dual  \eqref{15:sln} to the  latitude Wilson loop \eqref{eq:matrixfermionic} with $q=\cos\theta_0$. Pictures by E. Vescovi.
}
\label{15:fig}
\end{figure}

We start by noting that minimal surfaces dual to Wilson loops in $\mathbb R^3$ inside $\mathbb R^4$ of 4d SYM (with fixed position in the internal space)  are straightforwardly embedded inside $AdS_4$ and hence are also solutions for the 10d sigma model dual to ABJM theory. However, the difference between 
$S^5$ and $\mathbb {CP}^3$ implies that non-trivial profiles in internal space become more subtle. We discuss two examples, see Figure~\ref{15:fig}.

{\it Generalized cusp}:  after conformally mapping $\mathbb R^3$ to $ S^2\times\mathbb R$,  the piecewise linear Wilson line with a cusp shown in Figure~\ref{10:cusp}  is mapped to a pair of anti-parallel lines separated by an angle $\pi-\varphi$ along a great circle on the $S^2$. The  (non-susy) string dual to the static configuration on $S^2\times \mathbb R$ for arbitrary values of $\varphi,\theta$ coincides with the solution found in    \cite{Forini:2012bb} for 4d SYM. The embedding, in global AdS $(t,\rho,\vartheta,\psi)$, takes the form  \cite{Maldacena:1998im, Drukker:1999zq,Drukker:2007qr,Drukker:2011za}
\begin{equation}
t=\tau,
\quad
\rho=\rho(\sigma),
\quad
\vartheta=\frac\pi 2, 
\quad
\psi(\sigma)=\sigma, 
\quad
\theta_1=\theta_1(\sigma),
\quad
\alpha=\varphi_1=0,
\label{15:sln+B}
\end{equation}
with $\rho$, $\theta_1$ expressible analytically in terms of Jacobi elliptic functions. The AdS coordinate $\psi$ varies between $[\varphi/2,\pi-\varphi/2] $ and the $\mathbb{ CP}^3$ coordinate $\theta_1$ in $[-\theta/2, \theta/2]$. The leading order expression for $\Gamma_{1/2}$ in \eqref{10:Gammacusp} at strong coupling is obtained from the string on-shell action stripping away the temporal extension. Expanding around the straight line configuration $\theta=\varphi=0$ one finds \cite{Forini:2012bb}
\begin{equation}
\Gamma_{1/2}=\sqrt{2\lambda}\left[\frac1{4\pi}(\theta^2-\varphi^2)+{\cal O}((\theta,\varphi)^4)\right],
\end{equation}
which vanishes for the BPS configurations $\theta=\pm\varphi$ and perfectly matches \eqref{10:bremder} using \eqref{15:brem}.

\emph{Fermionic Latitude}: This is an adaptation to ABJM of a solution found in \cite{Drukker:2006ga} for 4d SYM, with non trivial profile in internal space. The string solution spans a disc at fixed time in global AdS and a cap bound by an $S^1\subset \mathbb {CP}^3$
\begin{equation}
t=0,\quad
\sinh\rho(\sigma)=\frac1{\sinh \sigma},
\quad
\vartheta=\frac\pi 2, 
\quad\psi(\tau)=\tau,\quad \sin\theta_1(\sigma)= \frac1{\cosh(\sigma_0+\sigma)},
\quad
\varphi_1(\tau)=\tau,\quad\alpha=0\,.
\label{15:sln}
\end{equation}
As we approach the AdS boundary $\sigma\to0$ the worldsheet describes a circle in internal space, 
$\sin\theta_1(\sigma)\to\sin\theta_0=1/{\cosh\sigma_0} $. As $\sigma\to\infty$ the worldsheet closes up, $\theta_1(\sigma)\to0$, resulting in a disk topology. A supersymmetry analysis shows that 4 out of the 24 supercharges are preserved, hence, the solution is 1/6 BPS for generic values of $\theta_0$. For $\sigma_0\to\infty$ the worldsheet is fixed in internal space, \eqref{15:sln} reduces to the $AdS_2\subset AdS_4$ geometry originally found in \cite{Berenstein:1998ij, Drukker:1999zq} and 
supersymmetry enhances to 1/2 BPS.

The identification of the dual Wilson loops was elucidated in \cite{Correa:2014aga}. The worldsheet \eqref{15:sln} 
describes the 1/6 BPS fermionic loop presented in Chapters~\ref{chapter:7} and~\ref{chapter:8}. 
Indeed, the $M^I_J$ matrix is reconstructed from the string endpoints in $\mathbb {CP}^3\subset \mathbb C^4$ in terms of four complex coordinates $z^I$. Exploiting the  ansatz $M^I_J   = \delta^I_J -2\dot{z}_J\dot{\bar{z}}^I/|\dot{z}|^2$ proposed in \cite{Rey:2008bh} one obtains \eqref{eq:matrixfermionic} with  $q=\cos\theta_0$. 
The leading contribution to the fermionic latitude in the fundamental representation at strong coupling arises from the on-shell worldsheet action, which after appropriate regularization gives
\begin{equation}
\label{WferGuille}
\langle W^\text{fer}(q) \rangle\sim e^{\pi\sqrt{2\lambda}\cos\theta_0}\,.
\end{equation}
This result coincides exactly with the expansion of \eqref{8:fermionicgenus0} at strong coupling. Moreover, it also provides a non-trivial check of the strong coupling expansion of the Bremsstrahlung function \eqref{11:dernu}
\begin{equation}\label{15:brem}
B_{1/2}=\frac{\sqrt{2\lambda}}{4\pi}+{\cal O}(1).
\end{equation}

\subsection{The unknown: Symmetry is not enough}

The above examples are the only Wilson loops for which the dual worldsheets is identified. All other classical string solutions are expected to be dual to Wilson loops with some $M^I_J$ and $\eta_I$ couplings preserving locally $U(1)\times SU(3)$, but  the exact form has never been worked out. So we really have a full holographic dictionary for Wilson loops that are globally BPS or break SUSY at points (cusps). 

Even less is know for the 1/6 BPS Wilson loops. Based on the $SU(2)\times SU(2)$ R-symmetry of the bosonic loops, it was suggested in \cite{Drukker:2008zx,Rey:2008bh}, that they should be dual to  the 1/2 BPS worldsheet `smeared' over $\mathbb {CP}^1\subset \mathbb {CP}^3$. This statement has not been precisely defined and in fact smearing only over one $\mathbb{CP}^1$ breaks the $\bZ_2$ symmetry between the two $SU(2)$ factors. It is even less clear how to represent all the 1/6 BPS fermionic Wilson loop interpolating between the bosonic and the 1/2 BPS loops found in \cite{Ouyang:2015bmy}. Could they be realized in terms of mixing boundary conditions as in \cite{Polchinski:2011im}?

Further questions arise for Wilson loops in high dimension representation, where in the context 
of ${\cal N}=4$ SYM in 4d the holographic duals are D3-branes, D5-branes, or ultimately 
``bubbling geometries'' 
\cite{Drukker:2005kx, Hartnoll:2006hr,Yamaguchi:2006tq,Yamaguchi:2006te}. 
The analog of this in 3d has also not been resolved.

A D6 brane solution which is 1/6 BPS was found in \cite{Drukker:2008zx}, but no 1/2 BPS analog is known. 
There is a 1/2 BPS D2-brane 
(or M2-brane, in the M-theory frame), but unlike the 1/2 BPS Wilson loop, it has a continuous modulus 
and was identified as the holographic dual of a vortex loop operator \cite{Drukker:2008jm}. The back-reaction 
of this brane on the geometry is known in terms of bubbling geometries \cite{Lunin:2007ab}, but only 
in cases preserving 16 supercharges, so for $k=1,2$.

\subsection{Acknowledgements}
I would like to thank J.~Aguilera-Damia, D.~Correa, M.~David, R.~de Le\'on Ard\'on, N.~Drukker, A.~Faraggi, V.~Giraldo-Rivera, L.~Pando-Zayas, V.~Rathee, D.~Trancanelli and E.~Vescovi for many illuminating and joyful discussions. Research supported by PIP-1109 CONICET, PUE-{\it B\'usqueda de nueva f\'\i sica} CONICET and UNLP X791.

\ifaffiliation
\subsection{Full affiliation}

Instituto de F\'\i sica de La Plata (IFLP) - CONICET \&
Departamento de F\'\i sica, Facultad de Ciencias Exactas, Universidad Nacional de La Plata C.C. 67, 1900 La Plata, Argentina
\fi

\ifaffiliation
\authors{Valentina Forini, City, University of London
and
Humboldt-Universit\"at zu Berlin\\*
Edoardo Vescovi, Imperial College London}
\else
\authors{Valentina Forini and Edoardo Vescovi}
\fi
\section{Holography for ABJM Wilson loops II: Quantum strings}
\label{chapter:12}

The minimal surfaces reviewed in Chapter~\ref{chapter:15} are the supergravity, saddle-point approximation of the open string partition function holographically dual to the Wilson loop expectation value. Quantum string corrections may be evaluated in a semiclassical fashion as an expansion in inverse powers of the effective string tension, in this case $T=2\sqrt{2\big(\lambda-\frac{1}{24}\big)}$. 
In  the planar limit $N,k\to \infty$ and $\lambda\equiv N/k$ finite, fermionic Wilson loops at strong coupling are then computed expanding perturbatively the path integral for a  free, fundamental type IIA  Green-Schwarz string in the $AdS_4\times\mathbb{CP}^3$ background  with Ramond-Ramond four-form and two-form fluxes  
\begin{gather}
\label{12:WZ}
\langle W_\square \rangle=
Z_{\text{string}}\equiv 
\int \mathcal{D}\delta X \mathcal{D}\Psi \, e^{-S^{\text{IIA}}[X_ \text{cl}+\delta X,\Psi]}
\overset{T\gg 1}{=}
e^{-T\,\Gamma^{(0)}[X_{\text{cl}}]-\Gamma^{(1)}[X_{\text{cl}}]-\frac{1}{T}\Gamma^{(2)}[X_{\text{cl}}]+\dots}\,,
\end{gather}
where  $X_\text{cl}$ is the classical solution, $\delta X$ denote the quantum fluctuations of the bosonic string coordinates, $\Psi$ stands for the 10d Majorana-Weyl spinors and $T\,\Gamma^{(0)}\equiv S_{cl}$ is the classical result, a suitably regularized area of the minimal surface. While the computational setup for the quantum correction in~\eqref{12:WZ}  is substantially the same as in the  $AdS_5\times S^5$ case, the absence of maximal supersymmetry in the $AdS_4\times \mathbb{CP}^3$ background makes the construction of the corresponding superstring action  non-trivial.  
Also,  the more complicated structure of the background fluxes results in considerably more involved  spectrum and interactions. In general, computing the one-loop fluctuation determinant $\Gamma^{(1)}$ and higher-order corrections presents all the subtleties 
inherent to semi-classical quantization of strings in AdS backgrounds with fluxes~\cite{Forini:2017ene}. 
 
     
\subsection{One-loop determinants}

Evaluating  $\Gamma^{(1)}$ requires only the quadratic part of the Lagrangian, and for the fermionic sector its structure is well-known in terms of the type IIA covariant derivative. The one-loop path integral is given by functional determinants of matrix 2d differential operators, whose coefficients have a complicated coordinate dependence. While the isometries of the classical backgrounds of interest here reduce the problem to one dimension, non-diagonal mass matrices may hinder the solution to the spectral problem.
This is the case for the generalized cusp $\Gamma(\varphi,\theta)$, reviewed in Chapter~\ref{chapter:10}, at finite $\varphi$ and $\theta$  (respectively, the geometric and internal angles deforming the straight line). Setting to zero one of the angles, say $\theta=0$, the mass matrices diagonalise and the bosonic part of the Lagrangian reduces to six massless and two massive scalars, while the fermionic part can be expanded in two massless and six massive 2-dimensional spinors. 
The resulting partition function comprises the kinetic operators of these scalars in the denominator and 
fermions in the numerator as~\cite{Forini:2012bb,Aguilera-Damia:2014bqa}
\begin{gather}
\label{12:det}
\Gamma^{(1)}=-\log\frac{\det^{2/2}\left(i{\tilde{\gamma}}^aD_a\right)\det^{3/2}\left(i\tilde{\gamma}^aD_a
-{\frac{1}{2}\varepsilon^{ab}\gamma_{ab}}\right)
\det^{3/2}\left(i\tilde{\gamma}^aD_a+{\frac{1}{2}\varepsilon^{ab}\gamma_{ab}}\right)}
{\det^{6/2}\left(-\nabla^2\right)\det^{1/2}\left(-\nabla^2+R^{(2)}+4\right)\det^{1/2}\left(-\nabla^2+2\right)}\,.
\end{gather}
The dependence on the angle $\varphi$ is via the metric of the classical string, which defines the Ricci curvature $R^{(2)}$, the spinor covariant derivative $D_a$, the scalar Laplacian $\nabla^2$ and distinguishes curved Dirac matrices $\tilde{\gamma}^a$ from the flat $\gamma^a$.  The differential operators in \eqref{12:det}
are complicated functions of~$\varphi$, and $\Gamma^{(1)}$  can only be given in an integral form. 
Analytic expressions can be  obtained expanding in  small $\varphi$ (or small $\theta$, if $\varphi$ is set to zero) with standard methods, and a special care for  boundary conditions of massless fermions~\cite{Aguilera-Damia:2014bqa}.  
The associated Bremsstrahlung functions---see \eqref{10:brem12} with $N_1=N_2$ and in planar limit---agree to one-loop order $B_{1/2}=B_{1/6}^\varphi=\frac{\sqrt{2\lambda}}{4\pi}-\frac{1}{4\pi^2}+{\cal O}(\lambda^{-1/2})$ and are consistent with the field-theory predictions in Chapter~\ref{chapter:11}.\footnote{The lack of a holographic dual of the bosonic Wilson cusp prevents a genuine computation of $B^{\theta}_{1/6}$ and a check of $B_{1/6}^{\varphi}=2B_{1/6}^{\theta}$.}  

The case of strong-coupling quantum corrections for smooth, supersymmetric Wilson loops is much more subtle. The one-loop string partition function for the 1/2 BPS circle~\cite{Kim:2012nd} disagrees with the matrix model, something attributed to unknown, overall numerical factors in the measure of the path integral. The latter are believed to cancel in the \emph{ratio} of partition functions for loops with the same topology. Indeed, the prediction for the latitude-to-circle  ratio of the matrix models in \eqref{eq:3:ABJ(M)_matrix_model} and \eqref{8:fermionicWL} has been matched.   For small latitude angle, this was obtained in~\cite{Aguilera-Damia:2018bam}   evaluating $\Gamma^{(1)}$ in a perturbative heat-kernel approach.  
For finite latitude  angles, phase-shift~\cite{Medina-Rincon:2019bcc} and Gel'fand-Yaglom~\cite{David:2019lhr} methods can be used, where a key point (developed first in~\cite{Cagnazzo:2017sny}) is how to maintain diffeomorphism invariance in the regularization procedure. It is customary to evaluate determinants on the curved geometry transforming it to the flat cylinder,  working namely with conformally rescaled operators, {\it e.g.}\ $ \widetilde{\mathcal{O}}=\Omega^2(\sigma)\,\mathcal{O}$ for Laplacians. This transformation is singular at $\sigma=\infty$ (tip of the worldsheet disk) and requires an IR cutoff which, to be diffeo-invariant, must necessarily depend on the latitude angle. The resulting determinants read then
\begin{equation}
\label{12:chain}
 \det\mathcal{O}=\left(\frac{\det\mathcal{O}}{\det\widetilde{\mathcal{O}}}\right)_\text{anomaly}\left(\frac{\det\widetilde{\mathcal{O}}}{\det\widetilde{\mathcal{O}}_\infty }\right)_\text{cylinder}\det\widetilde{\mathcal{O}}_\infty\,.
\end{equation}
Above, the first factor is the conformal anomaly, which cancels among all operators as it should in a consistent string theory. In the second factor, where $\widetilde{\mathcal{O}}_\infty$ denote the asymptotic (Klein-Gordon and Dirac) operators, the IR regulator eventually cancels off, but a finite residue remains in the third factor.
This ``IR anomaly'' and a special choice of boundary conditions for massless fermions are the non-trivial contributions ensuring agreement with the field-theory prediction~\cite{Medina-Rincon:2019bcc, David:2019lhr}.  

\subsection{Higher orders}
Beyond one-loop, the string action expanded near a classical background is formally non-renormalizable~\cite{Roiban:2007jf}. 
The most efficient setup to verify explicitly UV finiteness is an AdS light-cone gauge-fixing for the string action~\cite{Uvarov:2009nk}. This is used 
in~\cite{Bianchi:2014ada} for the evaluation of the light-like Wilson loop to two-loop order at strong coupling, which is the state of the art in  $AdS_4\times \mathbb{CP}^3$ sigma-model perturbation theory.   
In the light-like limit the Feynman diagrammatics simplifies and standard techniques allow the reduction to a basis of two scalar integrals. Dimensional regularization, together with powerful cancellations of logarithmic divergent integrals, leads to a finite $\Gamma^{(2)}$. The resulting
$
h(\lambda)=\sqrt{\frac{\lambda}{2}}-\frac{\log 2}{2\pi}-\frac{1}{48\sqrt{2\lambda}}+{\cal O}(\lambda^{-1})
$
matches the integrability prediction for the ABJM cusp anomaly~\eqref{13:hstrong} \cite{Gromov:2008qe,Gromov:2014eha}. 

\subsection{Future directions}
In the BPS cases, it would be important to have a better understanding of individual Wilson loops rather than their ratios, {\it e.g.}\ proving the triviality of the string one-loop partition function for the 1/6 BPS fermionic latitude~\cite{Cardinali:2012ru} recently considered in~\cite{Medina-Rincon:2019bcc}.  
A remarkable development for our understanding of the string path integral (both in the $AdS_4\times \mathbb{CP}^3$ and $AdS_5\times S^5$ backgrounds) would come from giving an intrinsic string-theory derivation of  the exact localization results, calculating the one-loop exact determinant for string fluctuations around the appropriate localization ``worldsheet locus''.   
In the non BPS case, data at finite coupling may be obtained 
with lattice field theory methods, as in~\cite{Bianchi:2016cyv}, discretizing the  Lagrangian of~\cite{Uvarov:2009nk} expanded around the chosen minimal surface 
and using Monte Carlo techniques.
Another stimulating direction, on the lines of~\cite{Giombi:2017cqn}, is to use the  Type IIA action expanded in fluctuations near the 1/2 BPS straight line ($AdS_2$) minimal surface to evaluate correlators of string excitations via  Witten diagrams.  This should give the strong coupling prediction  for the  correlators of elementary operator insertions on the Wilson line with protected scaling dimensions, see Chapter~\ref{chapter:9}, defining a defect CFT$_1$ living on the line. 

\subsection{Acknowledgements}
We thank L. Bianchi, M. S. Bianchi, A. Br{\'e}s, V. Giangreco M. Puletti and O. Ohlsson Sax for collaboration on \cite{Forini:2012bb,Bianchi:2014ada}. The works of VF is partially funded by the STFC grant ST/S005803/1 and by the Einstein Foundation Berlin.  The work of EV is funded by the STFC grant ST/P000762/1.

          
\ifaffiliation
\subsection{Full affiliation}
Department of Mathematics, City, University of London, Northampton Square, EC1V 0HB London, UK\\
Institut f\"ur Physik, Humboldt-Universit\"at zu Berlin, Zum Gro\ss en Windkanal 6, 12489 Berlin, DE\\
The Blackett Laboratory, Imperial College, London SW7 2AZ, UK
\fi

\authors{Fedor Levkovich-Maslyuk \ifaffiliation, \'Ecole Normale Sup\'erieure Paris and IITP Moscow\fi}
\section{Integrability I: The interpolating function $ h(\lambda)$}
\label{chapter:13}

\subsection{Background}
A key feature of the ABJM theory is integrability, {\it i.e.}\ an infinite-dimensional hidden symmetry which emerges in the 't~Hooft limit and leads to a plethora of nonperturbative results, especially for the spectrum of conformal dimensions/string states (see \cite{Klose:2010ki} for a review). This parallels a similar development for the 4d ${\cal N}=4$ super Yang-Mills theory \cite{Beisert:2010jr}. While this property remains a conjecture, it has been extensively tested for the case of equal gauge group ranks $N_1=N_2$ when the corresponding 't~Hooft couplings $\lambda_i=N_i/k$ become equal $\lambda_1=\lambda_2=\lambda$ (while $N_i$ and $k$ tend to infinity). We mostly focus on this regime. One major outcome of the integrability program is a finite set of functional equations, known as the Quantum Spectral Curve \cite{Gromov:2013pga}, which provide the exact spectrum of anomalous dimensions of all local single-trace operators \cite{Cavaglia:2014exa}. However, the result for ABJM is given in terms of an interpolating function $h(\lambda)$, which enters all integrability-based results (for example, the giant magnon dispersion relation \eqref{14:dispersion_relation} discussed in the next chapter) but itself is not fixed by integrability. 

\subsection{Fixing $h(\lambda)$}
Remarkably, one can make a proposal for the exact form of $h(\lambda)$ by relating an integrability calculation with the matrix model arising in the localization description of the 1/6 BPS Wilson loop presented in Chapters~\ref{chapter:3} and~\ref{chapter:4} \cite{Gromov:2014eha}. The idea behind this link comes from the observation that in ${\cal N}=4$ super Yang-Mills the expectation value of a circular Wilson loop is similar to the anomalous dimension of an operator built from $L$ scalars and $S$ covariant derivatives in the limit when $S\to 0$ \cite{Basso:2011rs}. It is natural to expect that some link of this type should also exist in ABJM theory. For ABJM one can use integrability to compute the anomalous dimension $\Delta$ of an operator with twist $L$ and spin $S$, when $S$ is small,
\begin{equation}
	\Delta=L+S+\gamma_L(\lambda)S+{\cal O}(S^2),
\end{equation}
where $\gamma_L(\lambda)$ is a nontrivial function of the coupling known as the slope function. In \cite{Gromov:2014eha} it was computed from the Quantum Spectral Curve analytically, and the result can be written concisely using building blocks defined as
\begin{equation}
\label{13:int1}
\oint {dy}{} \oint  {dz} \frac{\sqrt{y-e^{4\pi h}}\sqrt{y-e^{-4\pi h}}}{\sqrt{z-e^{4\pi h}}\sqrt{z-e^{-4\pi h}}},
\frac{ y^\alpha z^\beta}{z-y}
\end{equation}
with the integrals going around the cut $[e^{-4\pi h},e^{4\pi h}]$. As any integrability prediction, this result is written in terms of $h(\lambda)$. One can notice that the integrand here has four branch points, at 
\begin{equation}
\label{13:brpqsc}
	z_1=e^{4\pi h}\,,
	\qquad
	z_2=e^{-4\pi h}\,,
	\qquad
	z_3=\infty\,,
	\qquad
	z_4=0\,.
\end{equation}
Similarly, the integrand in the localization result of Chapter~\ref{chapter:4} for the 1/6 BPS Wilson loop also has four branch points, located at $a$, $1/a$, $b$ and $1/b$ in the notation of that chapter. Requiring that one set of four branch points can be mapped to the other one by a conformal transformation fixes $h$ in terms of $a$ and $b$, and gives as a result
\begin{equation}
\label{13:hab}
	h=\frac{1}{4\pi}\log\left(\frac{ab+1}{a+b}\right) \; .
\end{equation}
Using the explicit form of $a,b$ from equations \eqref{4:kappa}, \eqref{4:ab-kappa}, this gives
\begin{equation}
\label{13:exacth}
	\lambda = \frac{\sinh(2\pi h)}{2\pi}{}_3F_2\left(\frac{1}{2},\frac{1}{2},\frac{1}{2};1,\frac{3}{2};
-\sinh^2(2\pi h)\right),
\end{equation}
which is an equation that determines $h$ as a function of the coupling $\lambda$.

\subsection{Tests of the conjecture}
While this proposal for the exact form of $h(\lambda)$ may seem rather bold, it has passed several highly nontrivial tests. Namely, it reproduces all known data at weak coupling (four-loops, {\it i.e.}\ the first two coefficients \cite{Minahan:2009aq,Minahan:2009wg}) and at strong coupling (the first two terms in the expansion \cite{	McLoughlin:2008he,Abbott:2010yb,LopezArcos:2012gb}), with the corresponding expansions being
\begin{align}
	h(\lambda)&=\lambda-\frac{\pi^2\lambda^3}{3}+{\cal O}(\lambda^5), \qquad\lambda\to 0, \\
	\label{13:hstrong}
	h(\lambda)&=\sqrt{\frac{1}{2}\left(\lambda-\frac{1}{24}\right)}-\frac{\log2}{2\pi}+{\cal O}(e^{-\pi\sqrt{8\lambda}}), \qquad\lambda\to\infty.
\end{align} 
Curiously, the 1/24 shift at strong coupling matches the anomalous AdS radius shift discussed in \cite{Bergman:2009zh,Drukker:2010nc}.

\subsection{Extension to $N_1\neq N_2$}
For the case of unequal gauge group ranks, tests of integrability have been much more scarce. Nevertheless, the very algebraic structure of the Quantum Spectral Curve makes it rather nontrivial to introduce a second coupling into the problem, and it was conjectured in \cite{Cavaglia:2016ide} that the integrability description remains the same as for $N_1=N_2$, provided one uses a new function $h(\lambda_1,\lambda_2)$ instead of $h(\lambda)$. The above calculation then provides the same result \eqref{13:hab} for $h(\lambda_1,\lambda_2)$ where $a,b$ are still the branch points in the localization approach which are now indirectly fixed in terms of $\lambda_1,\lambda_2$ \cite{Marino:2009jd,Drukker:2010nc}. Remarkably, this conjecture reproduces \cite{Cavaglia:2016ide} all known data from the literature: one new coefficient at weak coupling at 4 loops \cite{Minahan:2009aq,Minahan:2009wg}, the strong coupling behavior, a prediction \cite{Minahan:2010nn,Bianchi:2016rub} to all orders in $\lambda_2$ when $\lambda_1\to 0$, and the expected invariance under the Seiberg-like duality which replaces $(\lambda_1,\lambda_2)\to (2\lambda_2-\lambda_1+1,\lambda_1)$. If correct, the proposal means that all integrability-based results computed for the $\lambda_1=\lambda_2$ case immediately carry over to the case of general $\lambda_1,\lambda_2$ via replacing $h(\lambda)$ by the new function $h(\lambda_1,\lambda_2)$.

\subsection{Future directions}
While the conjectured form of $h(\lambda)$ has passed a variety of tests, it is obviously important to put it on firmer ground, especially in the case of $N_1\neq N_2$. At weak coupling this seems highly challenging, since new tests would involve a six-loop calculation. At strong coupling one may be able to compute in the dual string model the exponential instanton corrections indicated in \eqref{13:hstrong}. A more definitive verification would be to compute one and the same observable from both localization and integrability, a promising candidate being the Bremsstrahlung function (see Chapter~\ref{chapter:14}). On a more conceptual level, the calculation described here provides a curious and rare link between the integrability and localization approaches, whose implications should be understood more completely. A fascinating possibility is that it could open the way to extend integrability beyond the planar limit, using as inspiration the calculation above where the branch cuts in the two pictures map to each other. In the localization approach the cuts of the spectral curve become discretized at finite $N_1,N_2$ \cite{Herzog:2010hf}, leading one to speculate that the same should happen to the cuts appearing in the integrability framework. 

\subsection{Acknowledgements}
I thank A.~Cavagli\`a, N.~Gromov and G.~Sizov for discussions and collaboration on related subjects. My work is supported by Agence Nationale de la Recherche LabEx grant ENS-ICFP ANR-10-
LABX-0010/ANR-10-IDEX-0001-02 PSL.

\ifaffiliation
\subsection{Full affiliation}
Departement de Physique, \'Ecole Normale Sup\'erieure / PSL Research University, CNRS, 24 rue
Lhomond, 75005 Paris, France; Also at Institute for Information Transmission Problems, Moscow 127994, Russia
\fi

\authors{Diego H. Correa \ifaffiliation, Universidad Nacional de La Plata\fi}
\section{Integrability II: The question of the cusped Wilson loop}
\label{chapter:14}

\subsection{Background}
The cusp anomalous dimension in ${\cal N}=4$ super Yang-Mills can be studied using integrability \cite{Drukker:2012de,Correa:2012hh}. First, one can solve the spectrum of local operators on a Wilson line using an asymptotic Bethe ansatz. The spectrum is described in terms of magnons propagating in an open spin chain, in which the boundaries are associated with the Wilson line at each side of the local operator. The reflection matrix of magnons is determined to all-loop order using the symmetries common to the Wilson line and the operator used as the reference in the Bethe ansatz. Then, rotating the reflection matrix of one boundary, one gets the spectrum of operators inserted in a cusp. Finally, the solution to the Thermodynamics Bethe Ansatz (TBA) equations for the ground state, when the size of the insertion is shrunk to zero, gives the cusp anomalous dimension.

When we turn to ABJM theories, since matter fields are bifundamentals of $U(N_1)\times U(N_2)$, the spin chain describing the spectral problem is alternating, because distinct types of fields occupy odd and even sites. One can take for instance
\begin{equation}
(C_1\bar C^3)^\ell\,,
\end{equation}
as a reference state invariant under an $SU(2|2)\subset OSp(6|4)$ \cite{Gaiotto:2008cg}. Fundamental excitations can be of type $A$ or $B$ \cite{Minahan:2008hf}, depending if they propagate over odd or even sites of the chain. There is an additional symmetry $U(1)_{\sf extra}$, under which type $A$ and $B$ magnons have opposite charge, and magnons accommodate in a $(\textbf{2}|\textbf{2})_A\oplus(\textbf{2}|\textbf{2})_B$ representation of $SU(2|2)$ \cite{Gaiotto:2008cg}. The numerical values of the central extensions of the algebra $\su(2|2)$, that  characterize the representations, can be related to the energy  and momentum of the magnons. For short representations, a relation between them gives rise to a dispersion relation \cite{Beisert:2006qh}
\begin{equation}
E(p) = \tfrac{1}{2}\sqrt{Q^2 + 16 h^2(\lambda) \sin^2({p/2})}\,.
\label{14:dispersion_relation}
\end{equation}
This includes an unspecified function of the coupling which in ABJM is non trivial. The residual symmetry $SU(2|2)\times U(1)_{\sf extra}$ constrains the scattering of magnon excitations on the chain. The alternating nature of the spin chain splits the S-matrix into blocks, but the $U(1)_{\sf extra}$ implies that the type of magnon $A$/$B$ is preserved in the scattering. Then, symmetry fixes the $AA$, $BB$ and $AB$ scatterings  to the famous $SU(2|2)$ S-matrix \cite{Ahn:2008aa} known to specify an integrable bulk scattering problem.

The very first question to address in the hope that the cusp anomalous dimensions in ABJM theories could be computed using integrability is whether ABJM Wilson loops impose integrable open boundary conditions for insertions along the loop.
This spectral problem would be integrable if the open spin chain Hamiltonian for the mixing of the operator insertions could be diagonalized with a Bethe ansatz. For that it is necessary that the reflection matrix satisfies the Boundary Yang-Baxter Equation (BYBE). Since the ultimate goal would be to obtain all-loop expressions for the cusp anomalous dimension, one would need to determine all-loop expressions for the reflection matrix of magnon excitations. Thus, a way to proceed  is to use the symmetries common to the Wilson loop and the Bethe ansatz reference state to constrain the reflection matrix and see if the latter is consistent with the BYBE.

\subsection{Symmetries}
Common symmetries and supersymmetries between the Wilson line and the reference vacuum state depend on the relative orientations in the internal space and can be sought either for 1/6 or 1/2 BPS Wilson lines.

The bosonic 1/6 BPS Wilson line with $M^I_J = {\sf diag}(-1-1,1,1)$, as in \eqref{2:bosonic}, is invariant under supersymmetry transformations generated by $\bar\Theta^{12}_+$ and $\bar\Theta^{34}_-$. If one considers it as the boundary to an insertion $(C_1\bar C^3)^\ell$, the two $SU(2)$
of the vacuum symmetry $SU(2|2)$ are broken and only $\bar\Theta^{12}_+$ survives. Thus, the overall residual symmetry is in this case $U(1|1)\times U(1)_{\sf extra}$. The most general right reflection matrix would be in principle of the form
\begin{equation}
{\mathbf R} = \begin{pmatrix} 
R_{AA} & R_{AB}
\\
R_{BA} & R_{BB}
\end{pmatrix}.
\label{14:RR}
\end{equation}
However, since the $U(1)_{\sf extra}$ is preserved by the boundary, the mixing of different types of magnons is ruled out. Commutation of the action of ${\mathbf R}$ with the generators of the residual symmetry $U(1|1)$ restricts the form of the reflection matrix but leaves two undetermined functions in each block
\begin{equation}
R_{AA} = R_{AA}^0 \diag(1, r_{A}, e^{-ip/2},-r_{A}e^{ip/2}),
\qquad
R_{BB} = R_{BB}^0 \diag(1, r_{B}, e^{-ip/2},-r_{B}e^{ip/2}).
\label{14:16blocks}
\end{equation}
The BYBE, using \eqref{14:16blocks} and ABJM bulk S-matrix \cite{Ahn:2008aa}, would not be satisfied for generic undetermined functions unless they were further restricted to specific expressions.

The 1/2 BPS Wilson line seems more promising as one expects a larger residual symmetry. In the case it has $M^I_J = {\sf diag}(-1,1,1,1)$, it is invariant under transformations generated by $\bar\Theta^{1J}_+$ and $\bar\Theta^{IJ}_-$ with $I,J\neq 1$. Under  these supersymmetry transformations, the superconnection changes as a supercovariant derivative and ${\cal P} \exp(i\oint {\cal L})$ changes covariantly, so one needs an appropriate $U(N_1|N_2)$ local insertion ${\cal Y}$ to preserve some of the original supersymmetries of the Wilson line. For an insertion involving $C_1\bar C^3$ it is possible to preserve the supersymmetries generated by $\bar\Theta^{12}_+$ and $\bar\Theta^{14}_+$. These supersymmetries altogether with the surviving $SU(2)$ R-symmetry give rise to a residual $SU(2|1)$. The blocks in a reflection matrix of the form \eqref{14:RR} would be further constrained in this case
\begin{equation}
R_{AA} = R_{AA}^0 \text{diag}(1, 1, e^{-ip/2},-e^{ip/2})\,,\qquad
R_{AB} = R_{AB}^0 \text{diag}(1, 1, e^{-ip/2},-e^{ip/2})\,,
\label{14:12blocks}
\end{equation}
and similar expressions for $R_{BB}$ and $R_{BA}$. The BYBE would be satisfied in this case when $R_{AB}^0=R_{BA}^0=0$  and there is no mixing between type $A$ and $B$ magnons.

\subsection{Open problems}

The residual symmetry analysis presented here is not found in the literature, but various colleagues who have considered the problem in the past have arrived to similar conclusions\cite{unpublished}. In both cases discussed, the residual symmetry does not seem to be enough to indicate whether the problem is integrable or not, either because some functions in the reflection matrix are left undetermined or because the mixing between type $A$ and $B$ magnons is not ruled out. It might be useful to do a perturbative derivation of the 2-loop open spin chain Hamiltonian for the mixing of the operator insertions. If no mixing between type $A$ and $B$ magnons is observed, one might take this to hold for all-loops as a working assumption.

Even in that case, one still needs the overall dressing phase of the reflection matrix. To determine it, one should derive a boundary crossing condition and look for the appropriate solution. In particular, introducing cusp angles by rotating one of the reflection matrices and considering the leading Luscher correction in the weak coupling limit, one should be able to reproduce the perturbative result for the cusp anomalous dimensions computed in \cite{Griguolo:2012iq} and, for small cusp angles, the perturbative Bremmstrahlung function \eqref{10:B12perturbative}.

If the TBA program could be completed for ABJM Wilson loops and the Bremmstrahlung function could be computed exactly as in  ${\cal N}=4$ super Yang-Mills \cite{Gromov:2012eu}, the comparison with the localization results of Chapter~\ref{chapter:11}, for example \eqref{11:B12exact}, could provide another way of determining the function $h(\lambda)$ seen in Chapter~\ref{chapter:13}.

\subsection{Acknowledgements}
I would like to thank C. Ahn, T. Bargheer, N. Drukker, D. Fioravanti, V. Giraldo-Rivera, M. Leoni and G.~Silva for useful discussions on this problem. Work supported by PIP 0681, PUE-B\'usqueda de nueva f\'isica and UNLP X850.

\ifaffiliation
\subsection{Full affiliation}
Instituto de F\'isica de La Plata - CONICET $\&$ Departamento de F\'isica, Facultad de Ciencias Exactas,
Universidad Nacional de La Plata C.C. 67, 1900 La Plata, Argentina
\fi


\authors{}

\end{document}